\documentclass[preprint]{aastex}

\received{}
\revised{}
\accepted{}

\usepackage[dvips,width=6.5in,height=9in,top=1in,left=1in]{geometry}

\shortauthors{Boyer et al.}
\shorttitle{``{\it Spitzer} Observations of $\omega$ Cen''}

\begin{document}

\title{A \textit {Spitzer Space Telescope} Atlas of $\boldmath{\omega}$\,Centauri:
  The Stellar Population, Mass Loss, and the Intracluster Medium}

\author{
  Martha L. Boyer\altaffilmark{1}, Iain McDonald\altaffilmark{2},
  Jacco Th. van Loon\altaffilmark{2}, Charles E. Woodward\altaffilmark{1}, 
  Robert D. Gehrz\altaffilmark{1}, A. Evans\altaffilmark{2},
  A. K. Dupree\altaffilmark{3}}

\altaffiltext{1}{Department of Astronomy, School of Physics and Astronomy, 116
  Church Street, S.E., University of Minnesota, Minneapolis, MN
  55455 USA; mboyer@astro.umn.edu}
\altaffiltext{2}{Astrophysics Group, School of Physical \& Geographical
 Sciences, Keele University, Staffordshire ST5 5BG, UK} 
\altaffiltext{3}{Harvard-Smithsonian Center for Astrophysics, 60
  Garden Street, Cambridge, MA 02138 USA}

%\special{!userdict begin /bop-hook{gsave 340 50 translate 20 rotate
%    /Times-Roman findfont 30 scalefont setfont 0 0 moveto 0.9 setgray
%    (V10/20.August.2007) show grestore}def end}

\begin{abstract}
We present a {\it Spitzer Space Telescope} imaging survey of the most
massive Galactic globular cluster, $\omega$\,Centauri, and investigate
stellar mass loss at low metallicity and the
intracluster medium (ICM). The survey covers approximately 3.2$\times$
the cluster half-mass radius at 3.6, 4.5, 5.8, 8, and 24~\micron{},
resulting in a catalog of over 40,000 point-sources in the
cluster. Approximately 140 cluster members ranging 1.5 dex in
metallicity show a red excess at 24~\micron{}, indicative of
circumstellar dust. If all of the dusty sources are experiencing mass
loss, the cumulative rate of loss is estimated at 2.9 -- 4.2 $\times$
10$^{-7}~M_\odot$ yr$^{-1}$, 63\% -- 66\% of which is supplied by
three asymptotic giant branch stars at the tip of the Red Giant Branch
(RGB). There is little evidence for strong mass loss lower on the
RGB.  If this material had remained in the cluster center, its dust
component ($\gtrsim$ 1 $\times$ 10$^{-4}~M_\odot$) would be
detectable in our 24 and 70~\micron{} images. While no dust cloud
located at the center of $\omega$\,Cen is apparent, we do see four
regions of very faint, diffuse emission beyond two half-mass radii at
24~\micron{}.  It is unclear whether these dust clouds are foreground
emission or are associated with $\omega$\,Cen. In the latter case,
these clouds may be the ICM in the process of escaping from the
cluster.

\end{abstract}

\keywords{globular clusters: individual
  ($\omega$\,Centauri)---infrared: general---stars: AGB and
  post-AGB---stars: mass loss---stars: variables---ISM: evolution}

\clearpage
\vfill\eject 
\section{{\sc Introduction}}
\label{sec:intro}

Despite their old age and low metal content, Galactic globular
clusters (GCs) are now known to be dust factories \citep{ramdani01,
origlia02, evans03, boyer06, jaccoea06a, lebzelter06, ita07}. This has
important implications for our understanding of mass loss from
low-mass and metal-poor stars and for the replenishment of the
interstellar medium (ISM) within old, metal-poor stellar systems, such
as GCs, dwarf spheroidal (dSph) galaxies, and galactic halos.

All stars more massive than $\simeq~0.8~M_\odot$ lose a significant
fraction of their mass before they deflagrate or leave a compact
remnant \citep[see reviews by][]{iben83,chiosi86}. In the final
stages of evolution, most stars enter higher stages of nuclear
burning, and the products of this nucleosynthesis are transported to
the stellar surface through convection or rotation. The chemically
enriched mass that is recycled (with some delay) into the ISM has a
significant impact on any further star formation, and drives the
chemical evolution within galaxies. In particular, the enrichment of
the ISM by dust is of great importance, as grains play important roles
in many ISM processes, including the formation of molecular hydrogen
\citep{gould63} and planet formation. The winds of low and
intermediate-mass stars on the upper red giant and asymptotic giant
branches (RGB and AGB) and massive red supergiants are prolific dust
producers \citep{gehrz89}. The dust is an integral part of the driving
mechanism of the most massive winds \citep[e.g.,][]{gehrz71, jura85,
gail87}, and our understanding of red giant mass loss relies heavily
on the details of the dust condensation process
\citep[e.g.,][]{salpeter74,gauger90,hofner97} --- which is poorly
understood in particular at low metallicity \citep[for a review,
see][]{jacco06}.

GCs are exquisite laboratories in which to study stellar winds because
they represent the closest match to a single stellar population at
metallicities that span over two orders of magnitude down to $\lesssim$
1\% solar \citep[cf.][]{gratton04}. Mass loss remains important
throughout the history of a GC \citep{jaccomc07}; even in old GCs,
stars lose $\sim$30\% of their mass, most or all of which is removed
from the cluster via a host of mechanisms including ram-pressure
stripping during Galactic plane crossings. RGB mass loss may be
responsible for the blue horizontal branches observed in metal-poor
GCs \citep{rood73}, and evidence for mass loss is seen in the line
profiles of chromospherically active or pulsating red giants in GCs
\citep{dupree84,cacciari04,mcdonald07,meszaros08}.

There is mounting evidence for mass loss from GC red giants in
the form of dusty winds \citep{ramdani01, origlia02}. The first
spectroscopic identifications of silicate dust are in the tip-AGB star
V1 in the massive GC 47\,Tucanae \citep{jaccoea06a} and subsequent
identification of minerals in other AGB stars in 47\,Tuc with {\it
Spitzer} IRS observations \citep{lebzelter06}. With [Fe/H]~$=-0.7$,
47\,Tuc is a metal-rich GC, providing an environment similar to that
found, for instance, in the Small Magellanic Cloud. To probe uncharted
terrain, it is important to study dust in GCs at lower
metallicity. The surprising detection of both interstellar and
circumstellar dust and gas in the metal-poor GC M\,15
\citep{evans03,boyer06,jaccoea06b} demonstrates that dust forms even
in extremely metal-deficient environments.

\subsection{{\it $\omega$\,Centauri}}
\label{sec:ocen}

The most massive Galactic GC, $\omega$\,Centauri, contains enough
stars in the dusty wind phase for an empirical
reconstruction of the evolution of mass loss. Its gravitational well
\citep[$v_{\rm esc}~=~44$ km s$^{-1}$;][]{gnedin02} could retain
interstellar matter if winds are slow. $\omega$\,Cen is located
approximately 15\degr{} out of the Galactic plane at a distance of
$\approx$5 kpc \citep{lub02,vandeven06,reddening}, so it is possible
to probe the stellar population to the cluster core.

The bulk of the stars in $\omega$\,Cen have [Fe/H] $=-1.7$
\citep{norris96,smith00}, a metallicity much lower than the
environments in which circumstellar dust has traditionally been
studied. Complications arise because $\omega$\,Cen also harbors a
small fraction of stars up to an order of magnitude more metal-rich
\citep{lee99, pancino00, pancino02} and an intermediate population of
stars with an abnormally high helium content \citep{norris96,
norris04}. Red giants in $\omega$\,Cen display a range in surface
abundances, from oxygen-rich, M-type stars with titanium oxides, stars
enhanced in CH or in cyanide (CN), to genuine carbon stars with
molecular carbon \citep[C$_2$;][]{jacco07}.  These disparate
populations cause difficulties in stellar evolution studies of
$\omega$\,Cen, but since stars from the metal-rich and metal-poor
populations can be distinguished by means of optical spectroscopy,
$\omega$\,Cen presents us with a unique opportunity to study the dust
and mass loss within a single environment as a function of atmospheric
chemistry.

Consensus has yet to be reached regarding the formation history of the
different subpopulations \citep{sollima05,
stanford06,villanova07}. The spread and peculiarities in elemental
abundances in $\omega$\,Cen have been interpreted as evidence that
this GC may be the remnant nucleus of a tidally disrupted dSph
\citep{zinnecker88, freeman93}. If so, $\omega$\,Cen may offer us
insight into the mass loss and chemical enrichment in the much more
distant nucleated dSphs, and perhaps dSphs in general.

Here, we present NASA {\it Spitzer Space Telescope} \citep{werner04,
    gehrz07} images of $\omega$\,Cen and a catalog including the
    magnitudes of over 40,000 point-sources from 3.6 to
    70~\micron{}. The 24~\micron{} mosaic, which enables reliable
    detection of individual circumstellar dust shells near the peak
    wavelength of their spectral energy distributions (SEDs), provides
    the most dramatic improvement over previous data.  We make use of
    proper motion \citep{vanleeuwen00} and radial velocity
    \citep{jacco07} measurements to separate cluster members from the
    substantial foreground stellar and background galaxy populations.
    Optical spectroscopic information permits us to compare the
    contributions from stars of different composition and evolutionary
    status to their infrared (IR) emission and dust production. Our
    observations are discussed in \S\,\ref{sec:obs}, photometry and
    the $\omega$\,Cen point-source catalog are described in
    \S\,\ref{sec:phot}.  The color-magnitude diagrams (CMDs), stellar
    populations, mass loss, and the intracluster medium
    (ICM) dust are discussed in \S\,\ref{sec:disc}, and our
    conclusions are presented in \S\,\ref{sec:concl}.

\section{{\sc Observations}}
\label{sec:obs}

We used the NASA {\it Spitzer Space Telescope} to obtain image maps of
the GC $\omega$ Cen (Program Identification [PID] 20648, PI: van Loon)
with the Infrared Array Camera \citep[IRAC;][]{fazio04} and the
Multiband Imaging Photometer for {\it Spitzer}
\citep[MIPS;][]{rieke04}.  Observations are centered at R.A. =
13$^{\mbox{\scriptsize{h}}}$26$^{\mbox{\scriptsize{m}}}$45\fs9,
decl. = --47\degr28\arcmin37\farcs0 [J2000.0] and map the cluster to
at least 3.2 times the half-mass radius \citep[$r_{\rm hm}$ $=$
4.8\arcmin{};][]{harris96} at each wavelength. Observation details are
summarized in Table~\ref{tab:obs}.

Images centered at 3.6, 4.5, 5.8, and 8~\micron{} were obtained with
IRAC on 2006 March 26 UT.  We observed in high dynamic range mode,
enabling us to correct saturated sources in the long exposures (10.4
s) with their counterparts in the short exposures (0.4 s) while still
achieving a high signal-to-noise ratio for the fainter sources.
Images were taken at nine dither positions in a cycling pattern to
build redundancy against outliers and artefacts. The resulting image
map contains a total of 441 images and covers an area approximately
35\arcmin{} $\times$ 35\arcmin{} around the cluster center.

MIPS images at 24 and 70~\micron{} were obtained on 2006 February 22
UT and 2006 March 02 UT with seven scan legs of approximately
0.5\degr{} in length. The resulting mosaics combine 4,620 images for
each wavelength and cover a total area of approximately 35\arcmin{}
$\times$ 55\arcmin{}. We designed our MIPS observing campaign with the
intention of obtaining only 24~\micron{} data, and the MIPS 70 and
160~\micron{} data were obtained automatically \citep[see][]{rieke04}.
Due to the non-functioning half of the 70~\micron{} array, the
70~\micron data do not have complete coverage. The warm state of the
telescope during the observations saturated the 160~\micron{} array,
rendering the 160~\micron{} data useless. MIPS observation details are summarized
in Table~\ref{tab:obs}.

\begin{figure}[h!]
%\epsscale{1} \plotone{f1} 
\includegraphics[scale=1.0,angle=0]{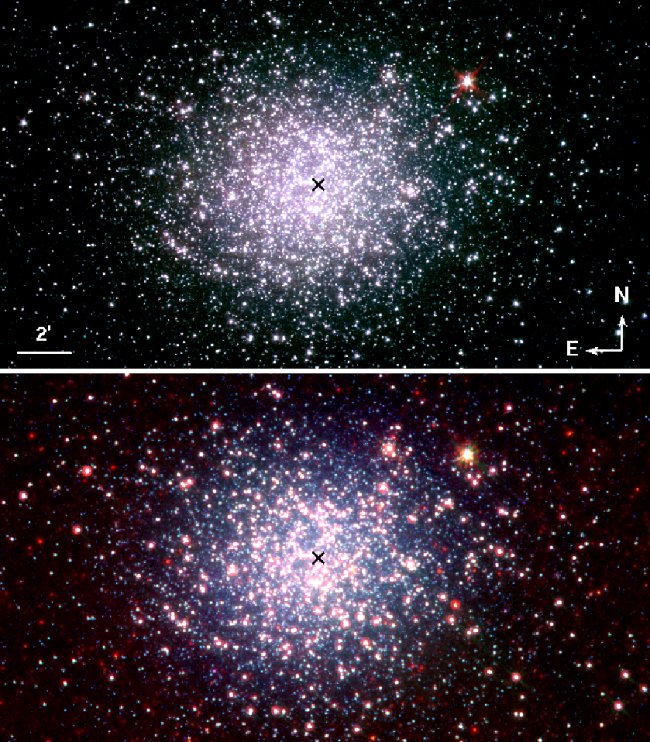} \figcaption{Three-color images
  of $\omega$ Cen. Upper panel: Blue is 3.6~\micron{}, green is
  4.5~\micron{} and red is 5.8~\micron{}. Lower Panel: Blue is
  3.6~\micron{}, green is 8~\micron{} and red is 24~\micron{}. The
  black ``X'' marks the cluster center. Dusty stars and/or background
  galaxies become visible at 8 and 24~\micron{} and appear red in the
  lower panel. No obvious intracluster medium dust clouds are
  visible. \label{fig:3color} }
\end{figure}

Raw MIPS and IRAC data were processed with the {\it Spitzer} Science
Center (SSC) pipeline, version
S13.2.0\footnote{http://ssc.spitzer.caltech.edu/archanaly/plhistory/}. The
Basic Calibrated Data (BCD) were post-processed with portions of the
2006 March 01 version of the SSC Legacy MOPEX software
\citep{makovoz05}. For IRAC, after applying an array distortion
correction, we implemented the MOPEX background-matching routine to
minimize pixel offsets in overlapping areas of the mosaics and the
MOPEX mosaicker to eliminate cosmic rays and other outliers. No
further cosmetic corrections were applied to the BCDs before
performing point-spread-function (PSF) photometry.  The final
mosaics have a pixel size of 1\farcs22~pix$^{-1}$. A three-color image
of the first three IRAC bands is presented in
Figure~\ref{fig:3color}a, in which very few sources appear with a red
or blue excess.

For the MIPS 24~\micron{} data, we created and applied
scan-mirror-dependent and scan-mirror-independent flat fields for all
images excluding the bias boost images, any image within 10\arcmin{}
of the cluster center, and any image containing a very bright
source. A 2-degree polynomial fit was applied over the entire image to
subtract the zodiacal background.  At 70~\micron{}, we first performed
pixel time-filtering and column filtering with SSC contributed
software from D. Fadda and
D. Frayer\footnote{http://ssc.spitzer.caltech.edu/mips/filter/}. We
then implemented the MOPEX background-matching and mosaic routines at
both 24 and 70~\micron{} before proceeding with point-source
extraction. An IRAC and MIPS three-color image is presented in
Figure~\ref{fig:3color}b, in which several cool dusty objects appear
green at 8~\micron{} and red at 24~\micron{}.

\section{{\sc Photometry}}
\label{sec:phot}

\subsection{{\it Point-Source Extraction}}
\label{sec:extraction}

Point-source extraction for 3.6 -- 24~\micron{} was done with the
DAOphot II photometry package \citep{stetson87}. PSFs were created
using a minimum of 20 isolated stars.  Sources brighter than 4
$\sigma$ over the background were chosen for extraction from the
mosaic image.  Any remaining outliers and other extended objects were
eliminated from the sample with a sharpness cut-off, which identifies
sources broader or narrower than the PSF. Photometric completeness at
each wavelength was determined through false star tests; 50\% and 90\%
completeness limits are listed in Table~\ref{tab:comp}.

\begin{figure}[h!]
%\epsscale{0.8} \plotone{f2} 
\includegraphics[scale=0.8,angle=90]{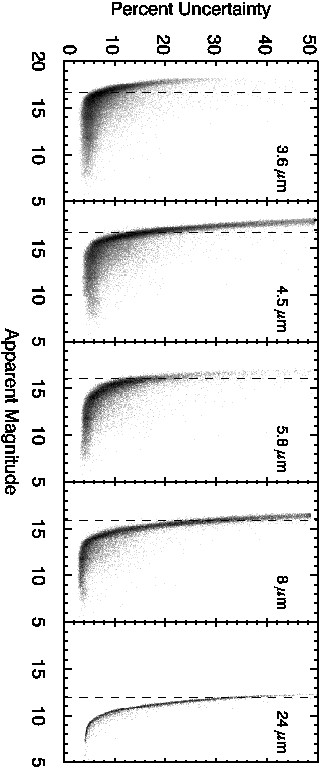} \figcaption{Photometric
  errors. The x-axis is binned by 0.1 magnitudes, and the y-axis is
  binned by 0.2 percentage points. The dashed lines mark the limit
  where photometry is 50\% complete (see Table.~\ref{tab:comp}). The
  errors for the bulk of the stellar population are low, but climb
  steadily at the faintest magnitudes. \label{fig:photerr} }
\end{figure}

The final fluxes and flux uncertainties are color-corrected according
to the IRAC and MIPS Data Handbooks versions 3.0 and 3.2.1,
respectively \citep{idh,mdh}, using a 5000~K blackbody, which is the
approximate temperature of a typical RGB star.  Fluxes in the
point-source catalog described in \S\,\ref{sec:catalog} have {\it not}
been corrected for reddening.  Where the reddening correction has been
applied for analysis of the data, $A_{\lambda}$ values are taken from
\citet{indebetouw05} and $E$($B-V$) $=$ 0.11 \citep{lub02}. The
pixel-phase-dependent photometric correction described by
\citet{reach05} was also applied to 3.6~\micron{} fluxes and their
uncertainties. This correction was applied only to 3.6~\micron{}
because it is not well defined for the other {\it Spitzer}
wavelengths. The flux uncertainties that we report in the {\it
Spitzer} catalog described in section \S\,\ref{sec:catalog} include
both the uncertainties quoted by DAOphot and the calibration
uncertainties specified by \citet{reach05} for IRAC and
\citet{engelbracht07} for 24~\micron{}.  All fluxes were converted to
stellar magnitudes using the zero fluxes reported in the {\it Spitzer}
Data Handbooks. The photometric uncertainty as a function of source
magnitude is given in Figure~\ref{fig:photerr}.

\subsection{{\it The Point-Source Catalog}}
\label{sec:catalog}

We have created a point-source catalog of $>$ 40,000 $\omega$\,Cen
objects detected at 3.6, 4.5, 5.8, 8, and 24~\micron{}.  An extraction
from this catalog is given in Table~\ref{tab:cat}, where column one is
the source number, column two is the source ID, columns 3 -- 7 are
{\it Spitzer} magnitudes, column 8 is the membership probability
\citep{vanleeuwen00}, columns 9 and 10 list
source type labels, and column 11 lists the Leiden identifiers (LEID)
from \citet{vanleeuwen00}. The source ID follows the standard {\it
Spitzer} naming convention, giving the truncated (J2000)
coordinates. The complete {\it Spitzer} catalog is available
electronically through the Centre Donn\'{e}es de Strasbourg (CDS).

Cluster members and non-members were identified by cross-correlating
the {\it Spitzer} source coordinates with those in the
\citet{vanleeuwen00} proper motion catalog. The \citet{vanleeuwen00}
catalog is not as deep as our survey nor is the coverage as wide, so
many {\it Spitzer} sources do not have proper motion
measurements. \citet{vanleeuwen00} derived positions and proper
motions by cross-referencing with the International Celestial
Reference Frame as defined by the Hipparcos catalog and extended to a
denser grid of stars by the Tycho-2 catalog. The coordinates of the
{\it Spitzer} sources are systematically offset from the
\citet{vanleeuwen00} sources, with mean offsets of $\Delta \alpha =
+0.40$\arcsec{} and $\Delta \delta =+0.25$\arcsec{} and standard
deviations of 0.15\arcsec{} and 0.10\arcsec, respectively. Coordinates
in the {\it Spitzer} catalog are not corrected for this offset.

If a source in any IRAC band was not associated with another IRAC,
MIPS, or \citet{vanleeuwen00} object, we considered it a false
detection and dropped it from the {\it Spitzer} catalog.  MIPS
24~\micron{} sources that do not have IRAC or optical counterparts remain in the
catalog since the very red SEDs of many MIPS sources peak in the
mid-IR and are thus fainter than the sensitivity limits in the bluer
{\it Spitzer} bandpasses.

Also included in the {\it Spitzer} catalog are notes in columns 9 and
10 indicating if the source is a galaxy, AGB, anomalous RGB (RGB-a),
or Horizontal Branch (HB) star, based on its locations on the CMDs
(see \S\,\ref{sec:cmd}).  The catalog also indicates sources that are
super-Ba- or CN-rich and/or are M-type (M), carbon (C), or post-AGB
stars (P), as identified by \citet{jacco07}.

\begin{figure}[h!]
%\epsscale{0.75}\plotone{f3}
\includegraphics[scale=0.75,angle=0]{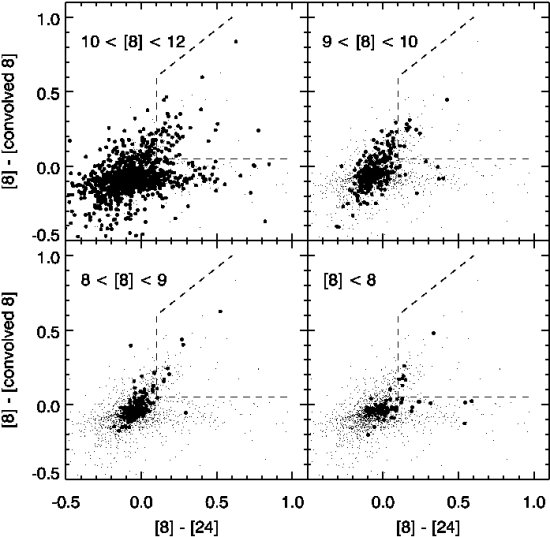} \figcaption{Color-color
  diagram showing which individual sources are affected by blending at
  24~\micron{}.  We convolved the 8~\micron{} image with the
  24~\micron{} PSF and extracted point sources from the resulting
  image. Each panel shows the entire population in small points, with
  the 8~\micron{} magnitude range indicated in each panel plotted in
  large, filled circles. Sources that have red [8] -- [convolved 8]
  colors are affected by blending, whereas sources with red [8] --
  [24] colors and [8] -- [convolved 8] $\approx$ 0 are truly red and
  likely harbor dust. We find that up to 25\% of the population with
  [8] -- [24] $>$ 0.1 is affected by blending. \label{fig:blend} }
\end{figure}

The differing resolution between IRAC and MIPS has caused some sources
that are isolated in IRAC to be blended in the 24~\micron{} mosaic, which
produces an artificially red [8] -- [24] color. To
determine which sources are potentially affected by blending, we convolved the
8~\micron{} mosaic with the 24~\micron{} PSF and identified sources
with elevated fluxes in the convolved mosaic compared to the unaltered
mosaic (Fig.~\ref{fig:blend}). These sources are marked with ``B24'', for ``blended at
24~\micron{}'', in column 10 of the {\it Spitzer} catalog. 

\subsection{{\it Luminosity Functions and Completeness}}
\label{sec:lumfunc}

\begin{figure}[h!]
%\epsscale{.9} \plotone{Figures/lumfunc_fin.ps}
\includegraphics[scale=0.65,angle=90]{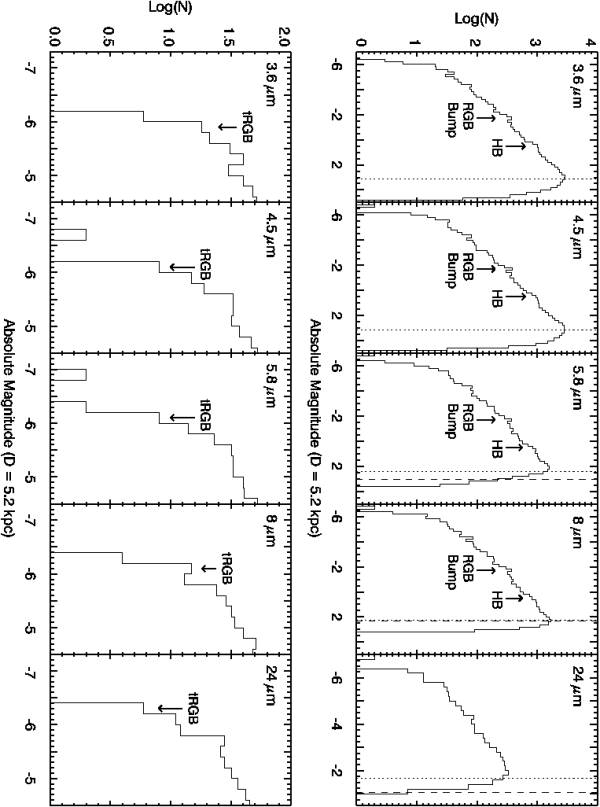} \figcaption{Luminosity
  functions for 3.6, 4.5, 5.8, 8, and 24~\micron.  The bin size is 0.2
  magnitudes and fluxes are corrected for reddening \citep[$E(B-V) =
    0.11$; ][]{lub02}.  In the upper panels, only sources detected in
  at least two bands are plotted and known non-members are
  excluded. In the lower panels, only known cluster members are
  plotted \citep{vanleeuwen00}. The dashed lines show the 3 $\sigma$
  sensitivity limit for each band (at 3.6 and 4.5~\micron{}, the
  sensitivity limit is fainter than the plot area shown), and the
  dotted lines mark the 50\% completeness limits. The photometry is
  nearly complete down to the 3 $\sigma$ sensitivity limit at 5.8 --
  24~\micron{}, but is less complete at the shortest wavelengths due
  to severe stellar crowding in the cluster core.  In the upper
  panels, the arrows indicate changes of slope that show the position
  of the RGB Bump/Clump and the HB. In the lower panels, the tRGB is
  marked as the point where the source counts drop by a factor of
  $\approx$4. \label{fig:lumfnc} }

\end{figure}

Luminosity functions for epoch two are displayed in
Figure~\ref{fig:lumfnc}. The distance modulus used to determine
absolute magnitudes is 13.58, given a distance of 5.2~kpc
\citep{reddening}. In the top panels, known non-members are excluded
and of the remaining sources, only those that are in at least two
adjacent wavebands are included. The IRAC luminosity functions show a
strong change in slope near an absolute magnitude of --2 that may be
the RGB bump or the RGB clump, both of which are located at absolute
magnitudes of $\approx$$-$2 -- 0 in the near-IR data of several GCs
\citep{valenti07}, depending on the metallicity of the cluster. The HB
is also visible at 3.6, 4.5, and 5.8~\micron{} just beyond an absolute
magnitude of zero as either a bump or change in slope.

The tip of the RGB (tRGB) is difficult to determine from the
luminosity functions, as the source counts are low at absolute
magnitudes brighter than $-$5.  The sparse population of AGB stars
above the tRGB is expected in GCs since this phase of stellar
evolution is particularly short-lived and also because the very low
mass of stars in GCs ($M_{\rm AGB}~\simeq~0.6 - 0.8~$$M_\odot$)
prevents AGB luminosity from climbing much above the tRGB.
Nevertheless, the source counts of known cluster members (lower panels
of Fig.~\ref{fig:lumfnc}) do drop by a factor of 3 to 4 past the bins,
which are 0.2 magnitudes wide, centered at absolute magnitudes of
$-5.9$, $-6.1$, $-6.1$, $-6.1$, and $-6.3$ at 3.6, 4.5, 5.8, 8, and
24~\micron{}, respectively. The K-band tRGB for $\omega$\,Cen is
located at $\approx$$-$6.04 $\pm$ 0.16 magnitudes
\citep{bellazzini04}. At 3.6~\micron{}, the tRGB location is $-$6
magnitudes for many stellar populations, including the Large
Magellanic Cloud \citep[LMC;][]{jacco05} and smaller stellar
populations such as the Local Group dwarf irregular galaxy WLM
\citep{jackson07}.

Photometric completeness in each waveband was determined by performing
false star tests, for which 2\% of the total stellar population that
mimicked the luminosity function was added to each mosaic at random
positions and PSF photometry was performed to attempt to recover these
false stars. The test was carried out 100 times for each
wavelength. Dotted lines mark the 50\% completeness limits in the
upper panels of Figure~\ref{fig:lumfnc}, which are listed in
Table~\ref{tab:comp}.  The dashed line in Figure~\ref{fig:lumfnc}
marks the sensitivity limit at each bandpass, and shows that the
photometry is the most incomplete at 3.6 and 4.5~\micron{}, which is
likely due to severe stellar crowding in the cluster core at these
wavelengths. Photometry at 5.8 -- 24~\micron{} is 50\% complete to
within two magnitudes of the 3 $\sigma$ sensitivity limits.

\section{{\sc Discussion}}
\label{sec:disc}

\subsection{{\it The Color-Magnitude Diagrams}}
\label{sec:cmd}

\begin{figure}[h!]
\begin{center}
%\epsscale{1} \plotone{Figures/cmds.ps} 
\includegraphics[scale=0.7,angle=90]{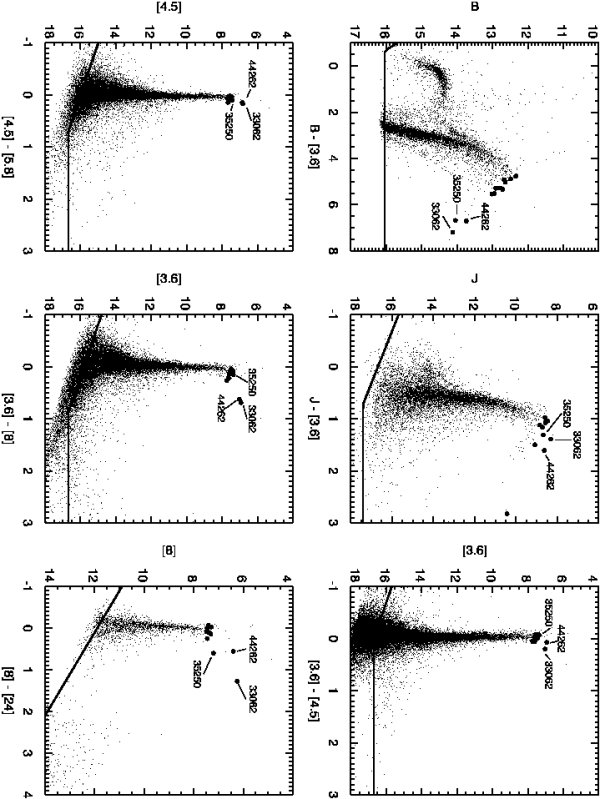} \figcaption{CMDs for $\omega$
  Cen.  Fluxes are corrected for reddening \citep[$E(B-V) = 0.11$;
  ][]{lub02}. $B$ magnitudes are from \citet{vanleeuwen00} and $J$
  magnitudes are from \citet{sollima04}. The solid lines mark where
  the photometry is 50\% complete in the case of {\it Spitzer} or mark
  the detection limit from \citet{vanleeuwen00} and
  \citet{sollima04}. The brightest confirmed cluster members at IRAC
  wavelengths are marked with large filled circles in each panel to
  show their evolution over color-magnitude space (non-members remain
  small dots). The three most extreme M-type stars are labeled with
  their LEID numbers. \label{fig:cmd1} }
\end{center}
\end{figure}

\begin{figure}[h!]
\begin{center}
%\epsscale{1} \plotone{Figures/cmds-color.ps} 
\includegraphics[scale=0.7,angle=90]{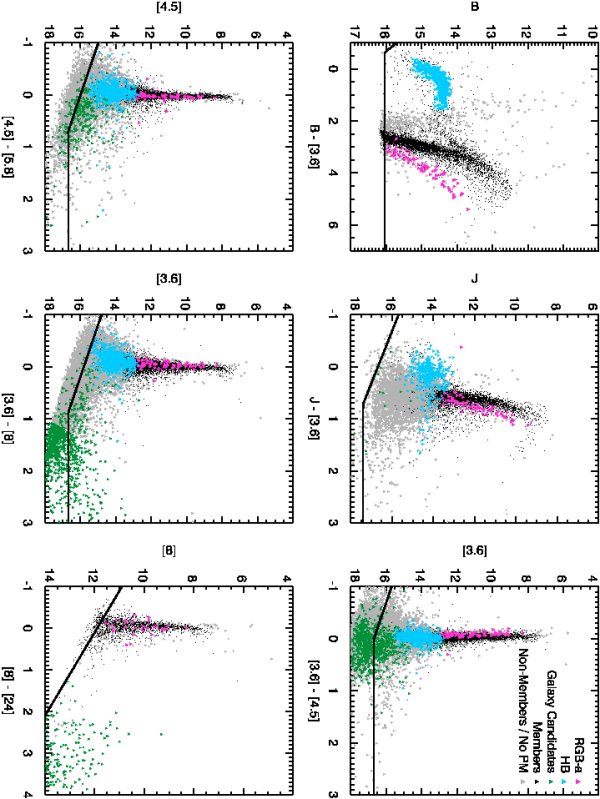} \figcaption{CMDs for $\omega$
  Cen. Same as Figure~\ref{fig:cmd1}, with different source types
  plotted in colors. Black points are confirmed cluster members
  \citep{vanleeuwen00}, gray points are not included in the
  \citet{vanleeuwen00} proper motion (PM) catalog, green points are
  possible background galaxies, blue points are HB stars, and magenta
  points are sources belonging to the RGB-a. HB and RGB-a stars were
  identified by their positions on the $B$ -- [3.6] CMD. At {\it
    Spitzer} wavelengths, the HB does not separate from the RGB as in
  optical CMDs, nor does the RGB-a except for a trend towards slightly
  blue [3.6] -- [4.5] colors. Background galaxies are identified by
  their [3.6] -- [8] and [8] -- [24] colors and appear very red at
  8~\micron{} due to PAH emission and at 24~\micron{} due to dust
  emission. \label{fig:cmd2} }
\end{center}
\end{figure}

Figure~\ref{fig:cmd1} displays the optical \citep{vanleeuwen00}, the
near-IR \citep{sollima04}, and our mid-IR CMDs. The CMDs are corrected
for reddening, as discussed in \S~\ref{sec:phot}, with $A_B$ values
from \citet{rieke85}, but the effect of reddening on the mid-IR
magnitudes is minimal.  The IRAC and MIPS CMDs shows very little
structure aside from the RGB, and at longer wavelengths, a very broad,
faint red plume.  Mid-IR isochrones that include all known RGB/AGB
properties, molecular bands, and dust considerations have yet to be
developed. For reference, the brightest IRAC sources are labeled with
large filled circles in Figure~\ref{fig:cmd1}, and the three brightest
M-type stars are also labeled with their LEID numbers.  Bright IRAC
sources that are not labeled are not cluster
members. Figure~\ref{fig:cmd2} shows the same CMDs with different
source types labeled in color. Black points are confirmed cluster
members (confirmed non-members are not plotted) and gray points, which
are mostly located at the faint end of the RGB, do not have matches in
the proper motion catalog.

Absorption in molecular bands may be responsible for some of the
spread towards red colors on the RGB, but cannot cause colors as red
as those of the dustiest stars. However, while many of the very red
stars on the mid- to lower-RGB in Figure~\ref{fig:cmd1} may be dusty,
it is very likely that some of the these sources are red due either to
mismatches across bands, photometric errors, or contamination from
cluster non-members.

The $B$ -- [3.6] CMD clearly displays the RGB, AGB, and HB. Also visible
as a sparsely populated branch to the red of the RGB is the RGB-a
(magenta points). The plume of stars that falls intermediate to the
RGB and the HB is mostly due to foreground contamination
\citep{bellazzini04}, however the brighest cluster members in this
plume ($B \lesssim$ 13, 1 $\gtrsim  B$ -- [3.6] $\gtrsim$ 3) include
post-AGB stars such as LEID 16018, 30020, 32015, 32029, and 43105. Note
that while \citet{vanleeuwen00} found that 16018 and 32029 are cluster
non-members, \citet{jacco07} determined through optical spectroscopy
that they do in fact belong to $\omega$\,Cen.  As such, their membership
probabilities have been set to 100\% in the {\it Spitzer} catalog.

 We identified HB stars (blue points) by their positions on the
 optical CMD \citep{vanleeuwen00}.  At IR wavelengths, the HB becomes indistinct from the
 RGB.  The stellar density profile shown in Figure~\ref{fig:denprof} shows a
paucity of HB stars in the center of the cluster. The non-central peak
of the HB density profile is explained by the dominance of the cluster
luminosity by bright AGB and RGB stars in the most crowded regions.

RGB-a stars are those located on the red branch of the RGB in optical
and near-IR CMDs \citep[see Fig. 5 of][]{sollima04}. The offset of the
RGB-a from the main RGB in optical and near-IR CMDs has been
attributed to a metallicity difference between the RGB-a and the
regular RGB \citep{lee99,pancino00,pancino02,origlia03}. Except for a
slight blue excess in the [3.6] -- [4.5] colors, no metallicity
separation is evident in the {\it Spitzer} CMDs.

\subsection{{\it The Asymptotic and Red Giant Branches}}
\label{sec:agb}

Inspection of the luminosity functions (Fig.~\ref{fig:lumfnc})
suggests that the stellar count begins to decrease at an accelerated
rate near an absolute magnitude of $-$5 at each wavelength. We chose
this magnitude as the divide at all {\it Spitzer} wavelengths between
mass-losing AGB candidates and the rest of the stellar population. We
identify 75 mass-losing AGB candidates, several of which have a red
excess, although mass-loss can occur even without a dusty
circumstellar envelope \citep{mcdonald07}. The AGB population is
slightly more centrally concentrated than the total stellar population
(Fig.~\ref{fig:denprof}), which is probably due to decreasing
completeness of fainter sources towards the core.

The locations on the CMD of stars that are identified as CN- and
super-Ba-rich by optical spectroscopy \citep{jacco07} are marked in
Figure~\ref{fig:populations}a.  The term super-Ba-rich distinguishes
stars with enhanced Ba through pre-enrichment from those that are
enriched beyond this, probably due to internal mixing processes. The
super-Ba-rich stars dominate the stars brighter than [24] $\approx$
8.5, supporting the suggestion by \citet{jacco07} that these stars are
enriched in Barium due to third-dredge up, and therefore delineate the
thermal pulsating AGB.  The super-Ba-rich stars may also be enriched
in carbon, but not so much as to warrant their classification as
genuine carbon stars.

Figure~\ref{fig:indiv} shows the [8] -- [24] CMD for cluster members,
non-members, and sources without proper motion measurements. Sources
affected by blending at 24~\micron{} are not plotted. The location on
the CMD of the majority of sources without proper motion measurements
suggests that most are likely galaxies. Stars brighter than [24]
$\approx$ 10 with [8] -- [24] $>$ 0.1 likely harbor dust, while
sources fainter than [24] $\approx$ 10 are dusty if [8] -- [24]
$>$ 0.2. Also marked in Figure~\ref{fig:indiv} and listed in
Table~\ref{tab:type} are 16 stars of interest, identified by
\citet{jacco07} as six M-type stars (open squares), five carbon stars
that show C$_2$ bands (open triangles), and five post-AGB stars
(closed circles), including Fehrenbach's star \citep{fehrenbach62}
near [24] = 9.25.

\citet{jacco07} obtained optical spectra for five of the extremely red
faint cluster members ([24] $>$ 10 and [8] -- [24] $>$ 0.8); these stars
range from [Fe/H] $=$ $-$1.5 to [Fe/H] $=$ $-$2, and two of the stars
are abnormally hot ($T \approx$ 7000 K). The remaining three stars
appear to be normal RGB stars. Visual inspection suggests that none of
these five sources are affected by blending at 24~\micron{}, however,
it is possible that HB stars or RGB stars that have little or no
emission at 24~\micron{} are spatially coincident with background
galaxies, causing them to appear to have red colors. This may also be
the case for the other $\approx$10 faint, extremely red stars that are
matched to cluster members.

{\it Spitzer} studies of GCs that do not have proper motion or radial
velocity unambiguous confirmation of membership will suffer from
contamination from red field stars mistaken for dusty cluster
stars. Many of the bright 24~\micron{} sources in our images are field
stars, including a variable M-type star with a period of 509 days
(LEID 34041). Proper motion measurements allowed us to identify almost
all bright and red cluster members in Figure~\ref{fig:indiv} against
field stars, with only one bright star remaining for which no proper
motion information is available. This bright source is most likely
LEID 55017, which is a confirmed cluster non-member.  The proper
motion of LEID 55017 is large enough that the {\it Spitzer}
coordinates are $\approx$4\arcsec{} different from the
\citet{vanleeuwen00} coordinates.

It is expected that metal-rich stars form more dust in their winds
than their low metallicity counterparts due to the higher abundance of
condensable material. We have metallicity information from
\citet{jacco07} for 32 of the $\approx$140 dusty stars in
Figure~\ref{fig:indiv}. The brightest potentially dusty stars in the
cluster range $-$2.25 $<$ [Fe/H] $<$ $-$1.25, i.e.\ nearly the entire
breadth of metallicities encountered in $\omega$\,Cen.  In fact, of
this small sample, the reddest stars have [Fe/H] $\leq$ $-$1.5
(Fig.~\ref{fig:dusty}) and two of the three brightest dusty stars
(LEID 33062, 44262, 35250) have [Fe/H] $\leq$ -2.0, which suggests
that dust production is not inhibited at these low metallicities.

\subsection{{\it Mass-Loss Rates}}
\label{sec:mlr}

Since mass-loss rates ($\dot{M}$) scale directly with dust optical depth
($\tau$), all dusty stars have the potential to drive a wind
\citep{ivezic95}. However, the value for $\dot{M}$ derived from the
optical depth also scales with the square root of the luminosity, so
the brightest stars contribute a larger fraction of the cluster's
cumulative mass-loss rate than the fainter stars. Of the $\approx$140
red cluster members that remain after eliminating blends and non-members
such as LEID 34041, which, as a bright M-type star with a period of
509 days, would dominate the cumulative mass-loss rate of the cluster
if included.  We used the AGB models computed by \citet{groenewegen06}
to determine the mass-loss rates corresponding to the color of the red
member stars with 0.15 $\gtrsim$ [8] -- [24] $\gtrsim$
3.1. Luminosities were determined by scaling to the
\citet{groenewegen06} mass-loss tracks for AGB stars with temperatures
of 3297 and 3850~K and with silicate and AlO$_{\rm x}$ dust
compositions, however differing bolometric corrections for stars with
different temperatures and metallicities are not taken into
account. For reference, the colors corresponding to two sample optical
depths are marked on the CMD in Figure~\ref{fig:indiv} and listed in
Table~\ref{tab:tau}. The final mass-loss rates ($\dot{M}_{\rm Final}$)
were scaled according to the following relationship:

\begin{equation}
\left(\frac{\dot{M}_{\rm{Final}}}{\dot{M}_{\rm G}}\right) = \left(\frac{v}{10~\rm{km~s}^{-1}}\right) \left(\frac{0.005}{\psi}\right) \left(\frac{L}{3000~L_\odot}\right)^{\frac{1}{2}},
\end{equation}

\noindent where $\dot{M}_{\rm G}$ is the unscaled mass-loss rate from
  \citet{groenewegen06}, $v$ is the wind velocity, $\rm \psi$ is
  the dust-to-gas mass ratio ($\psi~\approx$ 0.005 at solar
  metallicity), and $L$ is the luminosity. Following \citet{jacco00} and
  \citet{marshall04}, the velocity scales approximately as:

\begin{equation}
\left(\frac{v}{10~\rm{km~s}^{-1}}\right) = \left(\frac{\psi}{0.005}\right)^\frac{1}{2} \left(\frac{L}{10,000~L_\odot}\right)^\frac{1}{4}.
\end{equation}

\noindent From equation (2), the wind velocities of even the strongest
mass-losing stars in $\omega$\,Cen are slow ($v\ <\ 2\ \rm{km\
s^{-1}}$). Winds this slow may have difficulty driving a massive wind;
it is possible that the wind is driven by some other mechanism and
that the dust is a by-product. \citet{mcdonald07} suggest wind
velocities of 5 -- 10~km~s$^{-1}$ for AGB stars in GCs; if wind
velocities are 5$\times$ higher, the mass-loss rates would increase by
the same factor.

The \citet{groenewegen06} models were calculated assuming spherical
circumstellar dust shells, a constant mass-loss rate, and photospheres
with solar metallicity. The last of these assumptions affects the
stellar colors, which we used to determine the mass-loss rates.  While
we do take metallicity into account in the dust-to-gas ratio,
increasing the mass-loss rates by a factor of seven, the metallicity
dependence is not yet well established. The mass-loss rates computed
using the \citet{groenewegen06} models should therefore be considered
a rough estimate. Bolometric luminosities of the $\omega$\,Cen sources
can be estimated through fitting the SED or spectrum of each star,
facilitating a better estimate of the real mass-loss rates
(I. McDonald et al., in preparation).

Assuming that all stars have the metallicity that comprises the
majority of the population and that $\psi$ scales with metallicity
\citep[][$\rm{[Fe/H]}$ = --1.7, $\psi$ $\sim$
10$^{-4}$]{jacco00,jacco06}, we find that the three brightest M-type
stars (Fig.~\ref{fig:indiv}) contribute a total of $\dot{M}$\,(gas $+$
dust) $=$ 1.8 -- 2.8 $\times$ 10$^{-7}$ $M_\odot$ yr$^{-1}$, depending
on the temperature and dust composition. The combined mass-loss rate
of these three stars is an order of magnitude less than the mass-loss
rate measured for V1 in 47 Tuc with IR spectroscopy
\citep{jaccoea06a}, suggesting either that we have underestimated the
mass-loss rates or that the variance of $\dot{M}$ with metallicity is not a
linear relation. Figure~\ref{fig:mlr} and Table~\ref{tab:mlr} shows
that the cumulative mass-loss rate ($\Sigma\,\dot{M}$) of the
$\approx$140 dusty stars together is less than twice the combined
mass-loss rate of the three brightest M-type stars (LEID 33062, 44262,
35250). However, the fainter stars may be contributing more to the
total mass-loss rate of the cluster if their dust-to-gas ratios are
smaller.

To compare $\Sigma\,\dot{M}$ to the number of dusty stars present at
each magnitude, we also plot the percentage of stars that are dusty
($f_{D}$) in Figure~\ref{fig:mlr}, binned by half magnitudes. $f_{D}$
remains roughly constant at 10\% -- 20\% for [24] $>$ 9.5, causing the
cumulative mass-loss rate to slowly increase at [24] $>$ 8.  The
increase in $\Sigma\,\dot{M}$ at [24] $\approx$ 11 is due to the very
red sources ([8] -- [24] $>$ 1.0) at these magnitudes. It is possible
that many of the faint stars with red excess are not truly dusty (see
\S\ref{sec:agb}).  Since the cumulative mass-loss rate does not
increase much over more than 2 magnitudes on the upper-RGB, it is
unlikely that even fainter stars, which are generally warmer and do
not pulsate strongly, contribute significantly to the dust-traced mass
loss.

In a {\it Spitzer} study of 47 Tuc, \citet{origlia07} find that red
stars occupy the CMD from the tRGB down to the HB and conclude that
mass-loss is occurring at a significant rate along the entire
RGB. However, at least 25\% of the red sources in $\omega$\,Cen have
proven to have red colors because of blending effects at 24~\micron{}
(see \S~\ref{sec:catalog}). Figure~\ref{fig:mlr} shows that these
blended sources can increase the cumulative mass-loss rate by up to
17\%.  In more crowded clusters like 47 Tuc, blending will affect an
even larger percentage of red stars.  Indeed, it is evident in Figure
2 of \citet{origlia07} that the CMD covering the central region of the
cluster is affected by blending, since the many red sources claimed to
exist between --2 $< M_{\rm{bolometric}}$ $<$ 0 have no equivalent
in the CMD for the outer regions of the cluster. By eliminating these
blends, we find that significant dusty mass loss occurs only at or
near the tRGB.  

Two previous mid-IR studies of GCs detected several very red, faint
sources \citep{boyer06,ita07}.  Both studies offer the possibility
that these sources are dusty, mass-losing stars. While the nature of
the sources has yet to be established, we find that after weeding out
pesky field stars and background galaxies via proper motion
information, very few of the confirmed $\omega$\,Cen cluster members
reside in this region of the CMD (Fig.~\ref{fig:indiv}), and
significant dusty mass loss occurs only near the very tip of the RGB
or AGB. In the case of NGC 362 \citep{ita07}, these faint, red sources
may be background AGB stars belonging to the Small Magellanic Cloud,
however in other GCs, these sources are more likely to be background
galaxies.

If the current population in $\omega$\,Cen is representative of the
cluster's population at all times, then the cluster has lost at least
1 -- 2 $M_\odot$ of material (1 -- 2 $\times$ 10$^{-4}~M_{\odot}$ of
dust) through stellar mass loss since its last Galactic plane
crossing, which occurred more than 3.4 $\times$ 10$^6$ years ago
\citep{tayler75}. \citet{jacco07} find a few post-AGB stars in
$\omega$\,Cen.  Assuming a post-AGB lifetime of several $10^4$ yr and
0.1 $M_\odot$ of mass shed on the AGB, this could have produced up to
100 $M_\odot$ of lost mass.  This comparison suggests that either
these AGB stars lose on average less than 0.1 $M_\odot$ on the AGB, or
that our mass-loss rate estimates are too low (for instance because
the wind velocity is actually higher and/or the dust-to-gas ratio is
lower than assumed here).

\subsection{{\it Background Galaxies}}
\label{sec:galaxies}

\begin{figure}[h!]
%\epsscale{0.65} \plotone{f7} 
\includegraphics[scale=0.65,angle=0]{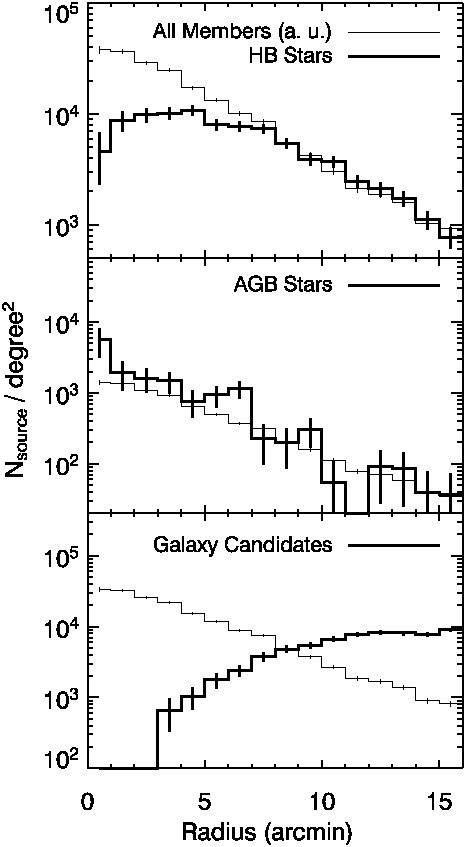} \figcaption{Source density
  profiles for $\omega$\,Cen. The thick line shows possible HB stars
  in the top panel, AGB stars in the middle panel, and galaxy
  candidates in the lower panel. For comparison, the density profile
  for all cluster members is shifted down to match each profile at $r
  =$ 8.5\arcmin{} and plotted in arbitrary units (a.u.)  as a thin
  line.  The AGB stars appear to be slightly more centrally
  concentrated than the general $\omega$\,Cen population, and there is
  a paucity of HB stars and galaxy candidates in the cluster
  center. The HB and AGB profiles might be explained by the lower and
  higher flux densities of HB and AGB stars, respectively, than the
  general stellar population rather than a real difference between the
  profiles. \label{fig:denprof} } \end{figure}

\begin{figure}[h!]
%\epsscale{1} \plotone{Figures/populations.ps} 
\includegraphics[scale=0.7,angle=90]{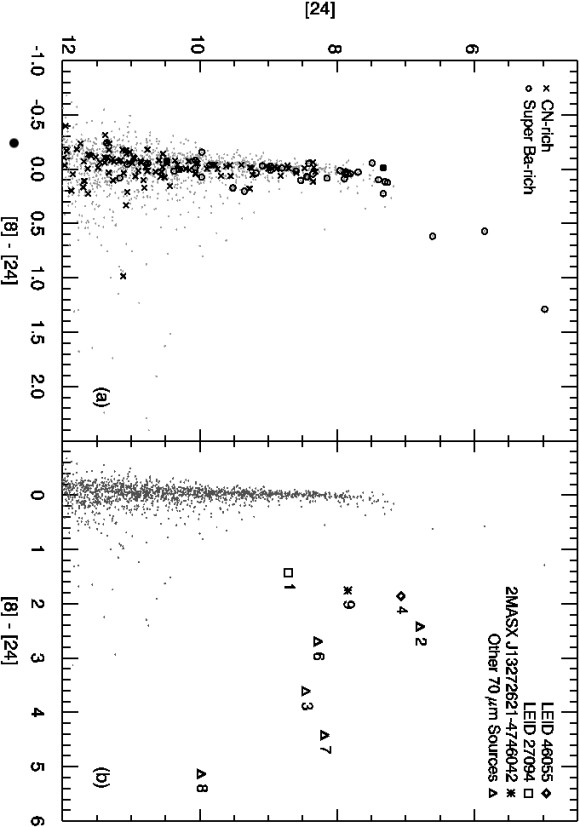} \figcaption{[8] -- [24] CMDs
  showing the location of various stellar populations.  Panel (a)
  shows super-Ba-rich and CN-rich stars as identified by their optical
  spectra \citep{jacco07}; super-Ba-rich stars delineate the upper
  part of the AGB possibly due to third dredge-up.  Panel (b) shows
  the 70~\micron{} sources.  Two of these are spatially coincident
  with cluster members (LEID sources), and one is a resolved spiral
  galaxy (2MASX J13272621--4746042). The numbers to the right of each
  symbol correspond to the source numbers in Tables~\ref{tab:70flux}
  and~\ref{tab:bb70}. All sources show excess 24~\micron{} emission.
   \label{fig:populations}}
\end{figure}

\begin{figure}[h!]
%\epsscale{1} \plotone{Figures/indiv.ps} 
\includegraphics[scale=0.65,angle=90]{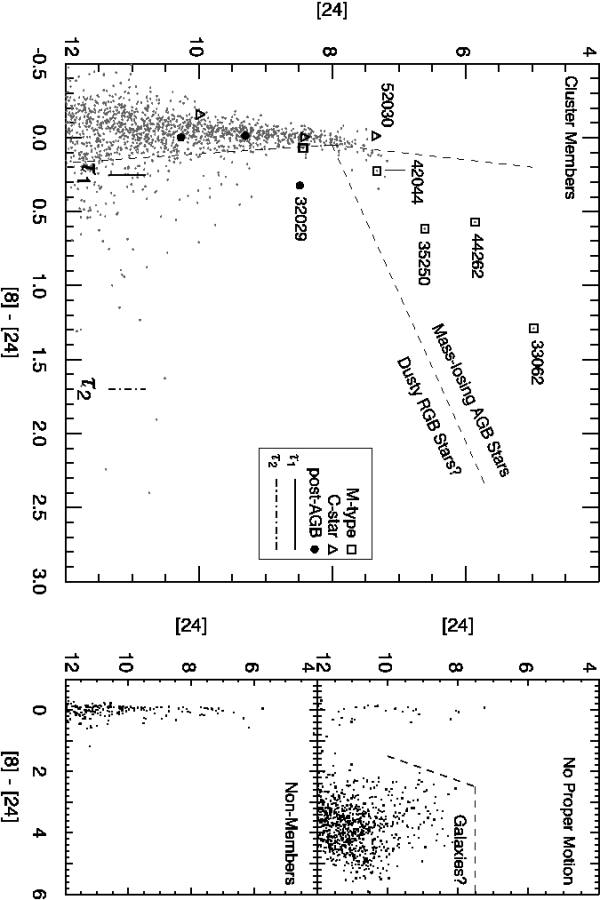} \figcaption{[8] -- [24] CMDs
  for $\omega$\,Cen.  Cluster members are plotted in the left panel,
  non-members are plotted in the lower right panel, and sources
  lacking proper motion measurements are plotted in the upper right
  panel. Sources that are potentially blended at 24~\micron{} (see
  \S\,\ref{sec:catalog}) are not plotted. Regions including AGB
  mass-losing stars and dusty RGB stars are separated by dashed lines
  in the left panel.  The colors corresponding to two sample optical
  depths ($\tau_1$ and $\tau_2$) are also marked on the left. See
  Table~\ref{tab:tau} for the values of $\tau$ for different dust
  compositions.  M-type stars, carbon stars, and post-AGB stars
  identified by their optical spectra \citep{jacco07} are also
  labeled, and the most extreme examples are also labeled with their
  LEID numbers.  Almost all bright and red cluster members in this CMD
  have been identified against field stars, with only one bright star
  remaining for which no proper motion information is available. This
  source may be LEID 55017, which has a large proper motion and is not
  a cluster member.  The bright luminosities and red colors of the
  three brightest M-type stars suggest that these three stars dominate
  the cluster mass-loss rate. All stars to the right of the vertical
  dashed line are included in the cumulative mass-loss rate
  (Fig.~\ref{fig:mlr}).\label{fig:indiv}}
\end{figure}

\begin{figure}[h!]
%\epsscale{1} \plotone{Figures/mlr_trend.ps} 
\includegraphics[scale=0.9,angle=0]{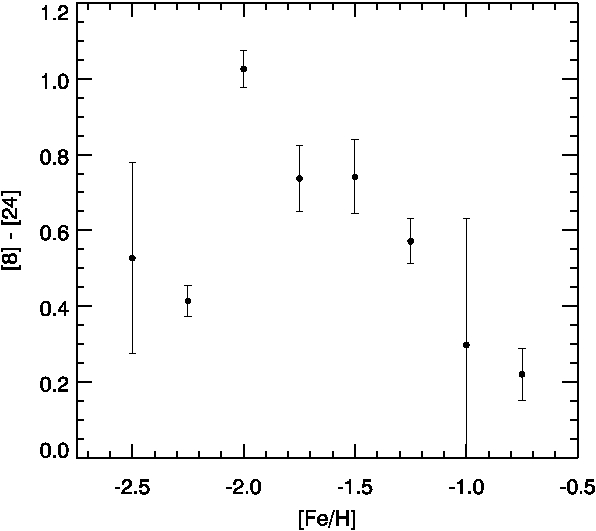} \figcaption{Metallicity
  versus average [8] -- [24] color. The color is a weighted average of
  all stars with a particular metallicity. Metallicities are
  determined by optical spectroscopy \citep{jacco07}. Dust production
  does not appear to be inhibited at low
  metallicity.\label{fig:dusty}}
\end{figure}

\begin{figure}[h!]
%\epsscale{1} \plotone{Figures/cum_mlr2.ps} 
\includegraphics[scale=0.65,angle=90]{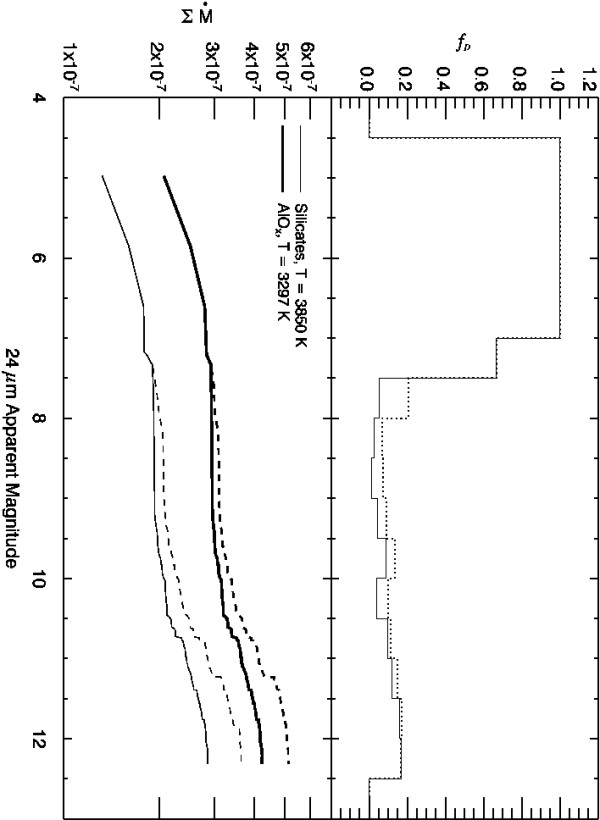} \figcaption{Cumulative
  mass-loss rate ($\Sigma~\dot{M}$) for $\omega$\,Cen. The upper panel
  shows the ratio of dusty to non-dusty stars ($f_{D}$).  The dotted
  line includes all observed members, while the solid line excludes
  potential source blends. The bin size is 0.5 magnitudes, but the
  large bin between 4.5 $<$ [24] $<$ 7 includes only the 3 brightest
  M-type stars, all of which are dusty. In the lower panel, individual
  mass-loss rates are derived from \citet{groenewegen06} models scaled
  to reflect the luminosity and metallicity dependence of the
  dust-to-gas ratio, and consequently, the wind speed. Sources to the
  right of the vertical dashed line in Figure~\ref{fig:indiv} are
  included in the summation and plotted with a solid line. The
  mass-loss rates are computed for four dust composition/temperature
  combinations, the most extreme of which are plotted; colder silicate
  and warmer AlO$_{\rm x}$ grains fall intermediate to the plotted
  lines. The dashed line shows the cumulative mass-loss rate obtained
  if potential blends are included, which can increase the total
  mass-loss rate by 12\% -- 17\% even in the relatively sparse
  environment of $\omega$\,Cen compared to other GCs. The sharp
  increases in $\Sigma~\dot{M}$ beyond [24] $\approx$ 10.5 are due to
  a few extremely red sources ([8] -- [24] $>$ 2). The total mass-loss
  rate for the cluster is 2.9 -- 4.2 $\times$ 10$^{-7}$ $M_\odot$
  yr$^{-1}$, 63\% -- 66\% of which comes solely from the three
  brightest M-type stars, depending on the dust temperature and
  composition.
\label{fig:mlr}}
\end{figure}

Background galaxies (green points in Fig.~\ref{fig:cmd1}) dominate the
faint, red plume in the [3.6] -- [8] CMD due to strong PAH emission at
8~\micron{} \citep{blum06,bolatto07}. The coverage overlap between the
3.6~\micron{} and 8~\micron{} mosaics is $\approx$0.3 deg$^2$,
yielding a density of 2823 galaxies/deg$^2$. This number increases to
3174 galaxies/deg$^2$ if we consider undetected galaxies behind the
cluster center (an area of $\approx$11 arcmin$^2$, see
Fig.~\ref{fig:denprof}). The red and faint region of the [3.6] -- [24]
CMD is also heavily populated with sources. If these
24~\micron{} sources are all background galaxies, the galaxy density
derived from this CMD is 2681 galaxies/deg$^2$, which increases to
3015 galaxies/deg$^2$ when considering sources located behind the
cluster core. When including all sources with very red colors and
faint magnitudes at all wavelengths, we find a total background galaxy
density of $\approx$1~$\times$~10$^{4}$ galaxies/deg$^2$
(Fig.~\ref{fig:denprof}).

To determine the expected background galaxy density, we queried the
SWIRE extragalactic database \citep{lonsdale04} for sources located in
the same region of the CMD as our galaxy candidates. In the six SWIRE
fields, there are $\approx$900 -- 2200 sources/deg$^2$ in the [3.6] --
[8] CMD and 900 -- 1900 sources/deg$^2$ in the [3.6] -- [24] CMD.  Our
images are slightly more sensitive than the SWIRE data, but when
comparing identical regions of the CMD, the galaxy densities in our
CMDs fall within the range of the SWIRE galaxy densities,
strengthening the argument that the sources we have identified as
galaxy candidates are indeed background galaxies. It is also possible
that the faintest sources in the red [3.6] -- [8] and [3.6] -- [24]
plumes are either stellar blends or regular RGB stars whose fluxes are
poorly determined due to their proximity to the completion limit,
causing them to have a non-zero color.

\subsection{{\it 70~\micron{} Sources}}
\label{sec:70micron}

\begin{figure}[h!]
%\epsscale{1} \plotone{f12} 
\includegraphics[scale=1.0,angle=0]{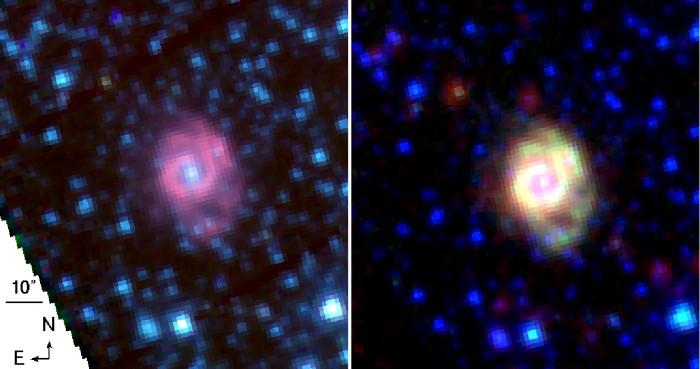} \figcaption{Three-color
  images of a resolved spiral galaxy (2MASX J13272621--4746042). {\it
    Left:} Blue is 3.6~\micron{}, green is 4.5~\micron{}, and red is
  8~\micron{}. {\it Right:} Blue is 3.6~\micron{}, green is
  8~\micron{}, and red is 24~\micron{}. 8~\micron{} PAH emission
  appears red in the left panel and green in the right panel. Dust
  appears red in the right panel. The SED of this galaxy
  (Fig.~\ref{fig:galsed}) resembles the SEDs of the other 70~\micron{}
  sources (Fig.~\ref{fig:70sed}).\label{fig:gal} } \end{figure}

Nine point-sources appear in the 70~\micron{} mosaic, although none
are the red, possibly dusty cluster members discussed in
\S\,\ref{sec:agb} and \S\,\ref{sec:mlr}. One of these sources is resolved
at the shorter wavelengths and appears to be a background spiral
galaxy (2MASX J13272621--4746042; Fig.~\ref{fig:gal}). This galaxy
shows a ring of PAH emission at 8~\micron{} that is reminiscent of the
famous star-forming ``Ring of Fire'', seen in M31
\citep{barmby06,gordon06}. Two of 70~\micron{} sources are near
confirmed cluster members \citep{vanleeuwen00,jacco07}; source 1 is
3.7\arcsec{} from LEID 27094, a distance that is within the
half-width, half-maximum of the 70~\micron{} PSF, and source 4 is
1.0\arcsec{} from LEID 46055.  Optical spectra for both LEID stars are
reported by \citet{jacco07} and reveal that both stars have [Fe/H]
$\simeq$ --1.25 and $T_{\rm eff}$ $\simeq$ 5000~K.  LEID 27094 is
CH strong and appears to be a normal RGB star, while LEID 46055 is CH
and CN strong, and is Ba-rich, suggesting that it may be enriched in
carbon from third dredge-up. Despite the proximity in position between
these two LEID sources and sources 1 and 4, the LEID sources may be
associated with warmer stars that appear in the IRAC mosaics as
secondary stars blended with sources 1 and 4, and not with the
corresponding 70~\micron{} source. The remaining six 70~\micron{}
sources are not near any known cluster members.

\begin{figure}[h!]
%\epsscale{1} \plotone{f13} 
\includegraphics[scale=.65,angle=0]{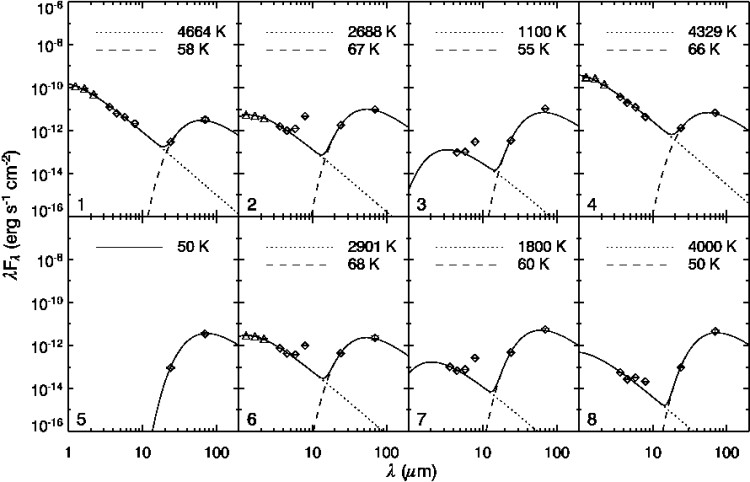} \figcaption{SEDs of
  70~\micron{} sources. Fluxes are corrected for reddening
  \citep[$E(B-V) = 0.11$; ][]{lub02}.  Sources 1 and 4 are spatially
  coincident with known cluster members and are potentially affected
  by blending between the known source and a fainter source that is
  the real counterpart of the 70~\micron{} emission. The 5.8 and/or
  8~\micron{} emission in sources 2, 3, 6, 7, and 8 lies significantly
  above the blackbody and is excluded from the fit.  Fitting results
  are summarized in Table~\ref{tab:bb70}. While the cold blackbody
  temperatures are consistent with a detached dust envelope such as
  those seen around post-AGB stars, they are also typical of the dust
  in star forming regions that dominate the galactic far-IR
  emission. The strong excess at 8~\micron{} and the slight excess at
  5.8~\micron{} could be due to either silicate emission or PAH
  emission. \label{fig:70sed} }
\end{figure}

We measured 70~\micron{} fluxes for the eight point-sources with the
aperture feature of {\it Imexamine} in version 2.12.2 of
IRAF\footnote{IRAF is distributed by the National Optical Astronomy
Observatories, which are operated by the Association of Universities
for Research in Astronomy, Inc., under cooperative agreement with the
National Science Foundation.}. Aperture radii were initially set to
3$\times$ the FWHM of each source and three iterations were performed
to find the optimal radius. The median of the background was
subtracted from the flux and the appropriate MIPS aperture correction
was applied. Fluxes are listed in Table~\ref{tab:70flux} and the 3.6
to 70~\micron{} SEDs are shown in Figure~\ref{fig:70sed}. Aside from
sources 1 and 4, all of the sources appear to be unaffected by blends
at all IRAC wavelengths. 

For each 70~\micron{} source, we performed a reduced-$\chi^2$ fit of a
two-component blackbody curve that include 2MASS $JHK$ fluxes to better
define the warmer blackbody. Source 8 falls off of our IRAC images, so
its IRAC fluxes were determined by using an earlier epoch of data from
the Gehrz Guaranteed Time Observing Program with {\it Spitzer} (PID
132) with data reduction identical to that described in
\S\,\ref{sec:obs}. The fits take the photometric errors into
account. Sources 3, 5, 7, and 8 are underconstrained for this fitting
technique when we exclude the anomalously bright 5.8 and 8~\micron{}
points from the fit; for these sources, we decreased the number of
fitting parameters by fixing the temperatures to 1000 to 14,000~K and
40 to 90~K for the blue and red blackbodies, respectively, and
choosing the temperatures that gave the best reduced-$\chi^2$. The
IRAC and MIPS color corrections were applied by iterating the fit
until the corrections remained constant within 1\%.  While a
dual-blackbody fits well with most of the 70~\micron{} sources, we
point out that the dust around these sources may be emitting over a
range of temperatures.

Better fits were obtained by excluding the 8~\micron{} point, which is
above the blue blackbody for all eight sources. To a lesser degree,
the 5.8~\micron{} point is also systematically brighter than the blue
blackbody component, suggesting that there is a range of dust
temperatures causing warm dust to radiate above the blackbody at
5.8~\micron{} and perhaps also at 8~\micron{}. However, the
5.8~\micron{} points are still included in the fit for source 3 in
order to keep the fit properly constrained. The resulting blackbodies
are overplotted in Figure~\ref{fig:70sed} and fitting results are
shown in Table~\ref{tab:bb70}.

\begin{figure}[h!]
%\epsscale{1} \plotone{f14} 
\includegraphics[scale=1.0,angle=0]{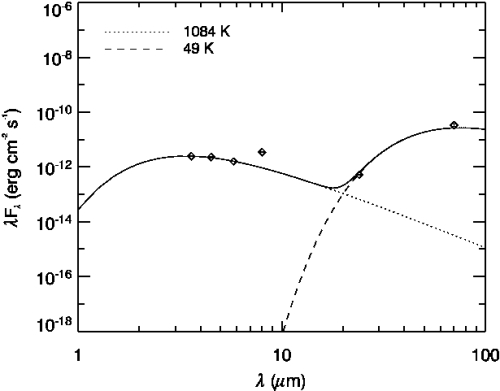} \figcaption{SED of galaxy
  2MASX J13272621--4746042. The 8~\micron{} excess is suggestive of
  PAH emission. Blackbody fitting results are summarized in
  Table~\ref{tab:bb70}. This SED resembles the SEDs of the other eight
  70~\micron{} sources, suggesting that they are also
  galaxies. \label{fig:galsed} } \end{figure}

The emission features at 5.8 and 8~\micron{}, where PAH emission
dominates, and the relatively flat SEDs of the non-blended sources
suggests that all or most of these sources are background galaxies
\citep[for examples of {\it Spitzer} galaxy SEDs, see][]{dale05}. For
comparison, an SED of the resolved spiral galaxy is shown in
Figure~\ref{fig:galsed}. We determined the flux by placing square
apertures of roughly 45\arcsec{} $\times$ 50\arcsec{} on the galaxy
and subtracting the median background. As with source 8, the galaxy
falls off of our IRAC coverage, so data from PID 132 is used to
determine its IRAC fluxes.  This SED is similar to the other eight,
except that it lacks the slight excess emission at 5.8~\micron{} that
many of the other sources exhibit.

The cold dust component of the blackbody fits for the eight
unidentified sources range in temperature from $\approx$50~K to 68~K
-- a temperature commensurate with these objects being post-thermal
pulse stars, stars suffering from episodic mass loss on the RGB, or
post-AGB stars. However, the temperatures of all eight sources are
also consistent with dust in star forming regions that dominate the
galactic far-IR emission. At a temperature of 48.8 $\pm$ 0.8~K, the
cold blackbody component of the resolved galaxy is colder than the
other eight 70~\micron{} sources, which may suggest that the warmer
sources are of a different nature. Although, with so few constraints
in the blackbody fits, one might argue that the temperatures of all
nine sources are consistent with each other.
 
If the 70~\micron{} sources are dusty AGB stars, the flux at
8~\micron{} may be included in the wing of a broad emission feature
centered at 9.5~\micron{} due to amorphous silicate grains that
overlaps with the IRAC 8~\micron{} bandpass. The 9.5~\micron{} feature
is also detected in the 8 -- 13~\micron{} spectrum of a particularly
red, large-amplitude variable star in 47~Tuc \citep{jaccoea06a,
lebzelter06}. The positions of the 70~\micron{} sources on the [8] --
[24] CMD are marked in Figure~\ref{fig:populations}b.  All sources lie
redward of the RGB and near the brighter end of the CMD. However, the
3.6 and 4.5~\micron{} magnitudes of sources 2, 3, 6, 7, and 8
(triangles) are fainter than 12.5 magnitudes, suggesting that these
sources are indeed galaxies. Sources 1 and 4 lie on the brighter end
of the RGB, but they are blended with the aforementioned LEID
sources at the shorter wavelengths, which will cause an artificial
increase in brightness.  While near-IR redshift measurements or radial
velocity measurements would conclusively determine the nature of these
sources, analysis of the {\it Spitzer} data strongly suggests that
they are all background galaxies.

\subsection{{\it The Intracluster Medium}}
\label{sec:icm}

Despite ongoing stellar mass loss in $\omega$\,Cen, there are no known
detections of any phase of ICM material in the cluster.  The wind
velocities discussed in \S\,\ref{sec:mlr} are $\ll$ 10 km s$^{-1}$ for
metal-poor, dust-driven winds from low-mass AGB stars. Such velocities
are much less than the escape velocity of $\omega$\,Cen at 1 r$_{\rm
hm}$ \citep[$v_{\rm esc}$ $=$ 44 km s$^{-1}$;][]{gnedin02}, but if the
dust and gas are not well coupled, then the dust velocity may be
higher. RGB and AGB wind velocities measured from H$\alpha$ emission
wings can be as high as the cluster escape velocity, however it is
likely that H$\alpha$ wings are due to pulsation and/or chromospheric
activity instead of the circumstellar shells of stars. Core
velocities originate in the chromospheres and are typically lower ($<$
15 km s$^{-1}$), but there have been at least two observations of
high velocities from the IR \ion{He}{1} line \citep{dupree92,
smith04}.

If the winds are indeed slow and the total mass-loss rate discussed in
\S\,\ref{sec:mlr} has been constant since the last Galactic plane
crossing, then, barring any ICM removal mechanisms and assuming that
all stars belong to the population with the lowest metallicity, 1 -- 2
$M_\odot$ of material should have accumulated in the cluster. This
mass estimate includes only dusty mass loss on the RGB and AGB; mass loss from
non-dusty stars on the RGB and from stars that lose significant mass
in such a brief period of time that it may easily have been missed
(e.g., associated with the formation of a planetary nebula) may cause
this estimate to increase by a factor of two or three.  Based on the
total amount of AGB mass loss, which is the only source of dust, we
expect a dust mass of $M_{\rm d}$ = $\psi\,M_{\rm gas}$ $\sim$ 1 -- 2
$\times$ 10$^{-4}$ $M_\odot$ of dust.  Since the estimate of the time
since the last Galactic plane crossing is a lower limit \citep[$>$ 3.4
$\times$ 10$^6$ yr;][]{tayler75}, the predicted ICM dust mass may be
higher.

The average 1 $\sigma$ sensitivity of our 24~\micron{} mosaic is
$\approx$2.5 $\times$ 10$^{-2}$ MJy/str. Following \citet{evans03},
the relationship between the dust mass and the expected
IR flux density is:

\begin{equation}
\frac{M_d}{M_\odot} = 4.79\times 10^{-17}\ f_\nu ({\rm mJy})\
\frac{D^{2}_{\rm kpc}}{\kappa_\nu B(\nu,T_d)},
\end{equation}

\noindent where $D_{\rm kpc}$ is the distance to $\omega$\,Cen,
  $\kappa_\nu$ is the dust absorption coefficient (cm$^2$ g$^{-1}$),
  $T_{d}$ is the dust temperature (K), and $B(\nu,T_{d}$) is
  the Planck function in cgs units. Assuming a standard
  Mathis-Rumpl-Nordsieck dust distribution \citep{mathis77} and an
  ISM-type dust composition of graphite and silicate grains,
  $\kappa_\nu$ is taken to be 56 $\pm$ 11 cm$^2$ g$^{-1}$
  \citep{ossenkopf94}, although this value could be up to a decade
  larger. A cloud with 1 -- 2 $\times$ 10$^{-4}$ $M_\odot$ of dust and
  an equilibrium temperature of 70~K would yield a 24~\micron{} flux
  density of 23 -- 45 mJy. With our sensitivity, we are capable of
  detecting 23 mJy to the 3 $\sigma$ level in our 24~\micron{} mosaic
  if all of the dust is uniformly distributed within an area of
  2\arcmin{} $\times$ 2\arcmin{}, or a radius of
  $\approx1/4~r_{\rm hm}$ A small, centrally located cloud
  at the 3 $\sigma$ level would be detectable over the
  background stellar emission in the 24 and 70~\micron{}
  mosaics. However, unlike in M15 \citep{boyer06}, our MIPS images do
  not show a large dust cloud located at the center of $\omega$\,Cen.

\begin{figure}[h!]
%\epsscale{0.9} \plotone{f15} 
\includegraphics[scale=0.9,angle=0]{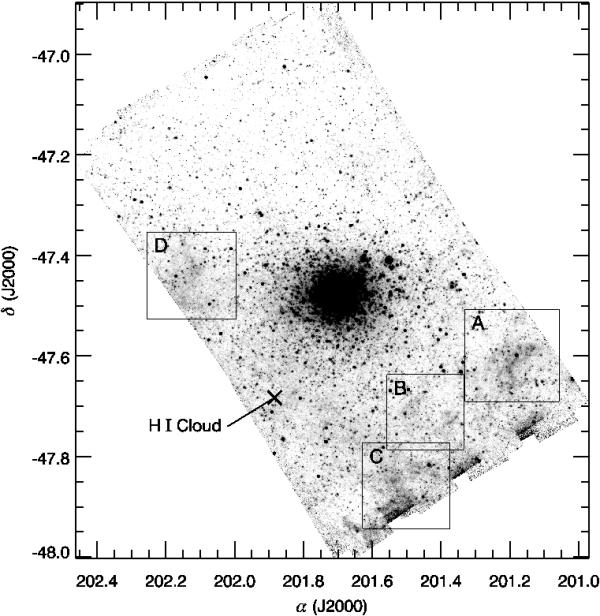} \figcaption{ 24~\micron{}
  image showing the locations of four regions of extended
  emission. The boxed regions are shown in Figure~\ref{fig:abcd}
  overlain with contours. Region A is shown in more detail in
  Figure~\ref{fig:icma}. An \ion{H}{1} feature is marked with an ``X''
  (Fig.~\ref{fig:hipass}). Low-level diffuse emission is also visible
  covering most of the image south of the cluster center. All dust
  clouds are outside of the cluster half-mass radius, suggesting that
  if they are part of the ICM, they may be in the process of leaving
  the cluster.\label{fig:icm}}
\end{figure}

\begin{figure}[h!]
%\epsscale{1} \plotone{Figures/abcd.ps}
\includegraphics[scale=.8,angle=90]{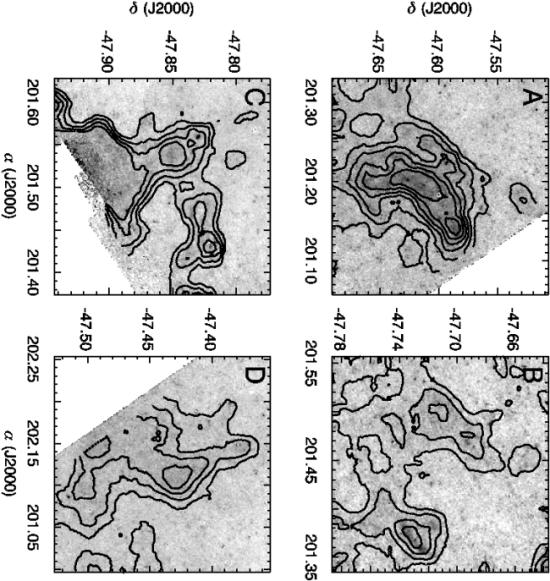} \figcaption{ Residual
  24~\micron{} image showing several regions of extended
  emission. Contours smoothed with a 25 pixel ($\approx$60\arcsec{})
  window are overlaid to show the morphology of each region. The
  bright region on the edge of the frame in panel C is a data
  artefact. \label{fig:abcd}}
\end{figure}

\begin{figure}[h!]
%\epsscale{1} \plotone{f17} 
\includegraphics[scale=1.0,angle=0]{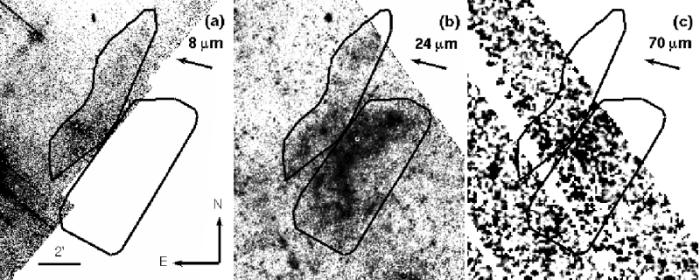} \figcaption{(a) Residual
  8~\micron{} mosaic. (b) Residual 24~\micron{} mosaic. (c)
  70~\micron{} mosaic. Each panel shows a close-up of possible ICM
  feature A located $\approx$20\arcmin{} from the cluster center. The
  arrows point towards the center of the cluster. The warm 8~\micron{}
  feature lies closer to the cluster center than the cooler feature at
  70~\micron{}. Both features appear at 24~\micron{}. This morphology
  suggests a possible association with the cluster. Note that the
  region of bright emission to the lower left of feature A at
  8~\micron{} is strongly affected by residual emission from an
  extremely bright source, and is not likely a real extended
  feature. \label{fig:icma} }
\end{figure}

While there is no central ICM dust cloud in $\omega$\,Cen, there is
faint, diffuse emission covering nearly the entire 24~\micron{} image
to the south of the cluster center, including several small regions of
concentrated extended emission on the outskirts of the cluster that
appear near the 3 $\sigma$ limit (Figs.~\ref{fig:icm}
and~\ref{fig:abcd}). Feature A (Fig.~\ref{fig:icma}) appears at 8, 24,
and 70~\micron{}, features B and C appear at 8 and 24~\micron{}, and
feature D appears only at 24~\micron{}, as the coverage is not
identical across all bands. The probability that these features are
artefacts is quite low as they each appear in more than 95\% of the
BCD images at various array positions.

It is uncertain whether features A -- D are associated with the
cluster or are part of the foreground. The morphology of feature A
provides the strongest evidence for cluster association, as the warmer
8~\micron{} emission lies closer to the cluster center than the cooler
24 and 70~\micron{} emission (Figs.~\ref{fig:abcd} and~\ref{fig:icma}).  Moreover, the
warmer portion is approximately perpendicular to the direction towards
the cluster center.  This configuration suggests the possibilities
that either a warm wind is pushing the ICM out of the cluster or that
the 8~\micron{} arc is optically thick and is being irradiated by the
cluster's stellar radiation field with cold dust in its shadow. A determination of the dust temperature would aid in ascertaining the
origin of feature A.  Unfortunately, only very small portions of
feature A are present in the {\it Spitzer} mosaics at all wavebands
(Fig.~\ref{fig:icma}), and the limited coverage makes it impossible to
reliably determine the temperature of the dust from an SED.

\begin{figure}[h!]
%\epsscale{1} \plotone{f18} 
\includegraphics[scale=1.0,angle=0]{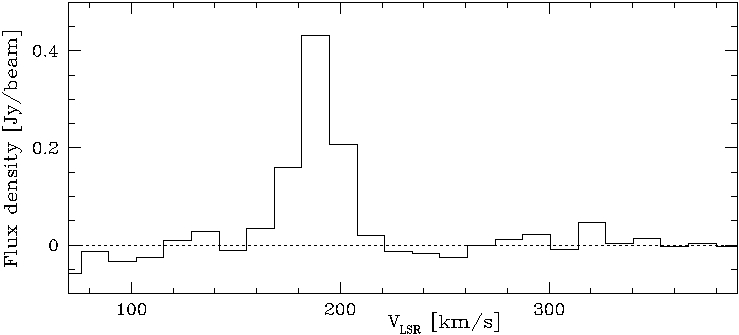} \figcaption{\ion{H}{1}
  detection near the velocity of $\omega$\,Cen ($v_{\rm
    LSR}~\sim~230~\rm km~s^{-1}$) in the HIPASS survey.  This feature
  is located $\approx$15\arcmin{} southeast of the cluster center (see
  Fig.~\ref{fig:icm}). The presence of \ion{H}{1} suggests that dust
  might also be present in the ICM of the cluster. \label{fig:hipass}
}
\end{figure}

The case for an ICM origin for features A -- D is strengthened by our
discovery of an unreported \ion{H}{1} cloud near the velocity of
$\omega$\,Cen (Fig.~\ref{fig:hipass}) in public data from the
\ion{H}{1} Parkes All-Sky Survey \citep[HIPASS; ][]{hipass}, which may
have also been detected by \citet{smith90} in earlier observations.
The center of this \ion{H}{1} feature is not spatially coincident with
any of the dust features, but rather lies $\approx$15\arcmin{} to the
southeast of the cluster center, just on the edge of the coverage of
our MIPS mosaic (Fig.~\ref{fig:icm}). The velocity of the \ion{H}{1}
feature ($v_{\rm LSR}~\sim$ 190 km s$^{-1}$) suggests a connection to
$\omega$\,Cen ($v_{\rm LSR}~\simeq$ 230 km s$^{-1}$). \citet{smith90}
argue instead that, because the feature is detected on only one side
of the cluster, it is more likely to be a small high-velocity cloud
that is part of the Magellanic Stream \citep[][$v_{\rm LSR}$
= 200 km s$^{-1}$]{bajaja85}.  While an association between the \ion{H}{1}
feature and the Magellanic Stream is a possibility, similar offset
\ion{H}{1} clouds have been discovered in larger dSph galaxies
\citep{bouchard03,bouchard06,young06}, so finding a cloud that is
associated with $\omega$\,Cen may not come as a surprise.  In M15,
dust and \ion{H}{1} trace each other in the ICM \citep{jaccoea06b}.
Therefore, if the \ion{H}{1} feature is indeed associated with
$\omega$\,Cen, it is reasonable to assume that dust is also present.
Since both the dust and the \ion{H}{1} feature are located well
outside 1 $r_{\rm hm}$, it is possible that we are seeing the ICM
in the process of escaping the cluster.

\section{{\sc Conclusions}}
\label{sec:concl}

Our {\it Spitzer} multi-wavelength and dual-epoch study of the most
massive Galactic globular cluster, $\omega$\,Cen, and the resulting
point-source catalog provide the most complete mid- to far-IR
atlas of any GC to date.  Despite $\omega$\,Cen's rather unusual
stellar population, the {\it Spitzer} CMDs show little structure beyond the division
between the RGB stars belonging to the cluster and background
galaxies. The HB stars and RGB-a stars are not separated from the RGB,
except for a slightly blue [3.6] -- [4.5] color for the RGB-a.

Nine sources are detected at 70~\micron{}, one of which is a resolved
spiral galaxy, and two of which are spatially coincident with optical
sources that are known cluster members. If emission at 5.8 and
8~\micron{} in the SEDs of the eight point sources is PAH emission,
then these sources are background galaxies. 

By cross-referencing with a catalog of optical spectra
\citep{jacco07}, we confirm that super-Ba-rich stars delineate the tip
of the AGB, likely due to a third dredge-up, and find that three
M-type stars not only have strong red excess, but are brighter than
the 24~\micron{} tRGB. Attributing the excess IR luminosity to dust
suggests that the three bright M-type stars dominate the cluster's
cumulative mass-loss rate (gas $+$ dust) of 2.9 -- 4.2 $\times$
10$^{-7}$ $M_\odot$ yr$^{-1}$. If the cluster mass-loss rate has
remained constant since the last Galactic plane crossing, then the
cluster has lost at least 1 -- 2 $M_\odot$ of dust-traced material
over the last 3.4 $\times$ 10$^{6}$ years. 

The dusty stars in our sample range over $\approx$1.5 dex in metallicity,
suggesting that dust production is not inhibited even in stars with
very low metal contents. In addition, our results show that, in
$\omega$\,Cen, significant dusty mass loss occurs near the very tip of
the RGB or AGB, concentrated in only a few individual stars, with
little evidence for such mass loss lower on the RGB.

While our images are sensitive enough to detect the amount of dust
expected in $\omega$\,Cen based on the current mass-loss rate, no
obvious ICM dust clouds are apparent in the cluster center at 24 or
70~\micron{}.  We do find several regions of faint extended
24~\micron{} emission on the outskirts of the cluster, but it is
unclear whether these clouds are associated with the ICM or are in the
foreground. If the dust features are associated with the cluster, they
are well outside the cluster half-mass radius, and could therefore be
ICM material that is in the process of escaping the cluster.

\acknowledgments

We thank the anonymous referee for his/her careful reading of the
manuscript and his/her valuable comments, which much improved the
presentation of these data. This publication makes use of data
products from the Two Micron All Sky Survey, which is a joint project
of the University of Massachusetts and the Infrared Processing and
Analysis Center/California Institute of Technology, funded by the
National Aeronautics and Space Administration and the National Science
Foundation. M.~.L.~B. is supported in part by the University of
Minnesota Louise T. Dosdall Fellowship.  M.~L.~B., C.~E.~W., and
R.~D.~G. are supported in part by NASA through \textit{Spitzer}
contracts 1276760, 1256406, and 1215746 issued by JPL/Caltech to the
University of Minnesota. I.~M. is supported by a STFC/PPARC
studentship. A.~K.~D. acknowledges research support from {\it Spitzer}
contract 1279224.

%%%%%%%%%%%%%%%%%%%%%%%%%%%%%%%%%%%%%%%%%%%%%%%%%%%%%%%%%%%%%%%
%%%% Bibliography
%%%%%%%%%%%%%%%%%%%%%%%%%%%%%%%%%%%%%%%%%%%%%%%%%%%%%%%%%%%%%%%%

\clearpage

%%%%%%%%%%%%%%%%%%%%%%%%%%%%%%%%%%%%%%%%%%%%%%%%%%%%%%%%%%%%%%%
%%%%% Figures 
%%%%%%%%%%%%%%%%%%%%%%%%%%%%%%%%%%%%%%%%%%%%%%%%%%%%%%%%%%%%%%%

% Fig 1: 3-color Image
\clearpage

% Fig 2: Photometry Errors
\clearpage

% Fig 3: Blending
\clearpage

% Fig 4: Luminosity Function
\clearpage

% Fig 5: CMDs for 3.6 microns
\clearpage

% Fig 6: Color CMDs for 3.6 microns
\clearpage

% Fig 7: Stellar Density Profile
\clearpage

% Fig 8: Populations CMD
\clearpage

% Fig 9: [8]-[24] CMD
\clearpage

% Fig 10: MLRs
\clearpage

% Fig 11: MLRs
\clearpage

% Fig 12: Resolved galaxy 3-color image
\clearpage

% Fig 13: 70 micron source SEDs
\clearpage

% Fig 14: Resolved galaxy SED
\clearpage

% Fig 15: Possible ICM in 24 micron map
\clearpage

% Fig 16: Possible ICM in 24 micron map
\clearpage

% Fig 17: Possible ICM
\clearpage

% Fig 18: HIPASS Feature
\clearpage

%%%%%%%%%%%%%%%%%%%%%%%%%%%%%%%%%%%%%%%%%%%%%%%%%%%%%%%%%%%%%%%
%%%%% Tables
%%%%%%%%%%%%%%%%%%%%%%%%%%%%%%%%%%%%%%%%%%%%%%%%%%%%%%%%%%%%%%%

%Table 1
\begin{deluxetable}{lccc}
\tablewidth{0pc}
\tabletypesize{\scriptsize}
\tablecolumns{6}
\tablecaption{{\sc Observational Summary} \label{tab:obs}}
\tablehead{\colhead{Parameter} &
  \colhead{IRAC} && \colhead{MIPS}}
\startdata

$\lambda$ (\micron) & 3.6, 4.5, 5.8 and 8 && 24 and 70 \\

Program ID & 20648 && 20648 \\

Date (UT)& 2006 Mar 26 && 2006 Feb 22 and Mar 02\\

AORkey   & 14955264 && (1495)5520, 5776, 6032, 6288,\\

& && 6544, and 6800\\

Coverage & 35\arcmin{} $\times$ 35\arcmin{} && 24~\micron{}: 35\arcmin{} $\times$ 55\arcmin{} \\

& && 70~\micron{}: 21\arcmin{} $\times$ 55\arcmin{}\\

Single Frame Exposure Time (s) & 10.4 && 24~\micron{}: 9.96\\

& && 70~\micron{}: 10.49 \\

Average Depth (no. frames) & 10 && 24~\micron{}: 67 \\

& && 70~\micron{}: 36 \\

Pixel Size (arcsec pixel$^{-1}$) & 1.22 && 24~\micron{}: 1.25 $\times$ 1.30\\

& && 70~\micron{}: 4.93 $\times$ 5.03\\

PSF size & 3.6~\micron{}: 1.7\arcsec{} $\times$ 1.6\arcsec{} && 24~\micron{}: 5.3\arcsec{} $\times$ 5.9\arcsec{}\\

& 4.5~\micron{}: 1.6\arcsec{} $\times$ 1.6\arcsec{} && \\
& 5.8~\micron{}: 1.6\arcsec{} $\times$ 1.7\arcsec{} && \\
& 8~\micron{}: 1.8\arcsec{} $\times$ 1.8\arcsec{} && \\

\enddata
\tablecomments{ \ Observations are centered at R.A. =
13$^{\mbox{\scriptsize{h}}}$26$^{\mbox{\scriptsize{m}}}$58\fs4,
decl. = --47\degr45\arcmin21\farcs8 [J2000.0]}
\end{deluxetable}

\clearpage

%Table 4
\begin{deluxetable}{ccc}
\tablewidth{0pc}
\tabletypesize{\small}
\tablecolumns{4}
\tablecaption{{\sc Completeness Limits for Different Cluster Radii} \label{tab:comp}}
\tablehead{\colhead{Wavelength}&\colhead{50\% Complete}&\colhead{90\% Complete}\\\colhead{(\micron{})}&\colhead{(mag)}&\colhead{(mag)}}
\startdata

3.6 & 16.7 $\pm$ 0.1& 15.9 $\pm$ 0.1\\
4.5 & 16.7 $\pm$ 0.1& 15.9 $\pm$ 0.1\\
5.8 & 16.0 $\pm$ 0.1& 14.5 $\pm$ 0.1\\
8   & 15.8 $\pm$ 0.1& 14.3 $\pm$ 0.1\\
24  & 11.9 $\pm$ 0.1& 11.6 $\pm$ 0.1\\

\enddata 
\tablecomments{ \ Completeness limits were determined from false star
  tests. These limits reflect the average over the entire cluster coverage;
  photometry is less complete in the core than on the outskirts of the
cluster.}
\end{deluxetable}

\clearpage

%Table 2
\begin{deluxetable}{rlrrrrrcllr}
\tablewidth{0pc}
\tabletypesize{\tiny}
%\rotate
\tablecolumns{14}
\tablecaption{{\sc Sample Source Catalog} \label{tab:cat}}
\tablehead{\colhead{No.} & \colhead{Source ID\tablenotemark{a}}&\multicolumn{5}{c}{Apparent Magnitude} &
  \colhead{Member-} & \colhead{Cand-} &
    \colhead{Note\tablenotemark{d}} & \colhead{LEID\tablenotemark{e}} \\ & &
    \colhead{3.6~\micron{}} & \colhead{4.5~\micron{}} & \colhead{5.8~\micron{}} & \colhead{8~\micron{}} &
    \colhead{24~\micron{}} & \colhead{ship (\%)\tablenotemark{b}} &\colhead{idate\tablenotemark{c}}& &\\
    \colhead{(1)}&\colhead{(2)}&\colhead{(3)}&\colhead{(4)}&\colhead{(5)}&\colhead{(6)}&\colhead{(7)}& \colhead{(8)} & \colhead{(9)}&\colhead{(10)}&\colhead{(11)}}

\startdata

18562&J132626.28$-$472520.06& 8.49(05)& 8.52(06)& 8.53(04)& 8.47(04)& 8.48(05)&    100&\nodata&     CN&  35090\\
18563&J132626.28$-$473229.59&14.04(09)&14.16(07)&14.24(07)&14.32(07)&  \nodata&\nodata&\nodata&\nodata&\nodata\\
18564&J132626.28$-$471144.58&16.58(08)&16.37(09)&15.86(13)&  \nodata&  \nodata&\nodata&\nodata&\nodata&\nodata\\
18565&J132626.28$-$474102.18&16.77(09)&16.68(16)&16.33(18)&  \nodata&  \nodata&\nodata&\nodata&\nodata&\nodata\\
18566&J132626.29$-$473737.68&17.51(12)&17.71(46)&  \nodata&  \nodata&  \nodata&\nodata&\nodata&\nodata&\nodata\\
18567&J132626.29$-$471200.23&17.20(11)&17.13(20)&  \nodata&  \nodata&  \nodata&\nodata&\nodata&\nodata&\nodata\\
18568&J132626.29$-$473049.11&13.59(05)&13.58(07)&13.68(06)&  \nodata&  \nodata&    100&     HB&\nodata&  47097\\
18569&J132626.29$-$472651.06&13.23(05)&13.20(06)&13.18(05)&13.16(10)&  \nodata&     98&\nodata&\nodata&  39111\\
18570&J132626.30$-$471915.97&17.04(09)&17.00(22)&  \nodata&  \nodata&  \nodata&\nodata&\nodata&\nodata&\nodata\\
18571&J132626.30$-$473515.89&17.30(14)&17.55(27)&  \nodata&  \nodata&  \nodata&\nodata&\nodata&\nodata&\nodata\\
18572&J132626.30$-$471416.24&16.00(06)&  \nodata&16.21(19)&  \nodata&  \nodata&\nodata&\nodata&\nodata&\nodata\\
18573&J132626.30$-$474159.18&13.22(06)&13.24(06)&13.28(04)&  \nodata&  \nodata&     86&\nodata&     CN&  72019\\
18574&J132626.31$-$471627.20& 9.34(04)& 9.28(06)& 9.31(05)& 9.29(04)& 9.30(06)&     18&\nodata&      P&  16018\\
18575&J132626.31$-$473111.71&14.94(08)&15.06(11)&  \nodata&  \nodata&  \nodata&\nodata&\nodata&\nodata&\nodata\\
18576&J132626.31$-$472320.08&13.98(06)&13.99(05)&14.00(05)&14.02(06)&  \nodata&\nodata&\nodata&\nodata&\nodata\\
18577&J132626.31$-$474246.15&17.80(20)&17.95(52)&  \nodata&16.08(35)&  \nodata&\nodata& Galaxy&\nodata&\nodata\\
18578&J132626.32$-$473543.47&13.12(04)&13.18(09)&13.11(08)&13.18(08)&  \nodata&\nodata&\nodata&\nodata&\nodata\\
18579&J132626.32$-$472534.32&10.13(05)&10.15(06)&10.15(05)&10.16(04)&10.18(09)&    100&\nodata&\nodata&  36061\\
18580&J132626.32$-$474718.12&  \nodata&15.72(08)&  \nodata&15.67(26)&  \nodata&\nodata&\nodata&\nodata&\nodata\\
18581&J132626.33$-$473132.57&12.23(08)&12.28(07)&12.24(06)&12.02(06)&  \nodata&    100&  RGB-a&\nodata&  49082\\

\enddata 

\tablecomments{ \ Magnitude uncertainties are quoted in
  parentheses. For example, a magnitude of 8.49 $\pm$ 0.05 is listed
  as 8.49(05).}  \tablenotetext{a}{ \ To save space, the Source ID is
  truncated to show only the source coordinates.  The full electronic
  catalog, available through CDS, includes the {\it Spitzer} prefix
  (SSTOCEN).}  \tablenotetext{b}{ \ Membership likelihoods are quoted
  as a percentage from \citet{vanleeuwen00}.}  \tablenotetext{c}{ \
  Candidate AGB, RGB-a, and HB stars and candidate galaxies are
  identified by their positions on optical and {\it Spitzer} CMDs.}
  \tablenotetext{d}{ \ The Note column specifies whether a source is
  CN- or super-Ba-rich and/or is a M-type (M), carbon (C), or post-AGB (P)
  star, as indicated by optical spectra \citep{jacco07}. The note also
  indicates if a source is potentially affected by blending at
  24~\micron{} (B24, see \S\,\ref{sec:catalog}).}  \tablenotetext{e}{ \ The LEID
  identification numbers are assigned to {\it Spitzer} sources via
  cross-identification with \citet{vanleeuwen00}.  }
\end{deluxetable}

\clearpage

%Table 5
\begin{deluxetable}{llrrrrrl}
\tablewidth{0pc}
\tabletypesize{\scriptsize}
\tablecolumns{8}
\tablecaption{{\sc {\it Spitzer} Magnitudes of M-Type, Carbon, and post-AGB Stars identified by \citet{jacco07}}\label{tab:type}}
\tablehead{\colhead{LEID} & \colhead{Source ID\tablenotemark{a}} & \colhead{m$_{3.6}$} &
  \colhead{m$_{4.5}$} & \colhead{m$_{5.8}$} & \colhead{m$_{8}$} &
  \colhead{m$_{24}$} & \colhead{Source Type}}
\startdata

 14043&J132750.37$-$471542.79& 11.75(06)&   \nodata& 11.77(04)&   \nodata&   \nodata&Carbon Star\\
 16018\tablenotemark{b}&J132626.31$-$471627.20&  9.34(04)&  9.28(06)&  9.31(05)&  9.29(04)&  9.30(06)&Post-AGB\\
 30020&J132549.95$-$472259.76& 12.62(05)& 12.62(06)& 12.61(06)& 12.66(04)&   \nodata&Post-AGB\\
 32015&J132538.44$-$472401.79& 10.26(05)& 10.23(07)& 10.22(07)& 10.27(04)& 10.27(09)&Post-AGB\\
 32029&J132605.17$-$472342.43&  9.11(05)&  8.96(05)&  8.82(04)&  8.81(04)&  8.48(05)&Post-AGB\\
 32059&J132628.13$-$472340.56&  8.45(05)&  8.52(06)&  8.52(04)&  8.40(05)&  8.40(05)&Carbon Star\\
 33062&J132630.19$-$472427.89&  6.96(05)&  6.88(06)&  6.74(05)&  6.27(04)&  4.98(05)& M-Type\\
 35094&J132628.81$-$472523.60&  8.43(05)&  8.56(07)&  8.48(05)&  8.50(05)&  8.43(05)& M-Type\\
 35250&J132737.72$-$472517.37&  7.39(05)&  7.45(06)&  7.35(05)&  7.23(04)&  6.61(05)& M-Type\\
 35252&J132738.29$-$472505.60& 14.21(04)& 14.07(10)& 14.49(08)& 14.15(10)&   \nodata& M-Type\\
 41071&J132614.42$-$472805.39&  9.82(04)&  9.93(06)&  9.89(05)&  9.81(04)&  9.97(08)&Carbon Star\\
 42044&J132605.35$-$472820.80&  7.59(04)&  7.74(06)&  7.67(04)&  7.55(05)&  7.33(05)& M-Type\\
 43105&J132627.22$-$472847.47&  9.56(04)&  9.54(06)&  9.46(04)&  9.44(04)&   \nodata&Post-AGB\\
 44262&J132646.37$-$472930.35&  7.05(05)&  6.85(05)&  6.68(04)&  6.42(04)&  5.85(05)& M-Type\\
 44484&J132726.37$-$472916.93&  8.44(06)&  8.63(05)&  8.53(05)&  8.51(05)&  8.44(05)& M-Type\\
 52030&J132601.59$-$473306.08&  7.40(05)&  7.48(06)&  7.44(05)&  7.32(04)&  7.33(05)&Carbon Star\\
 53019&J132544.03$-$473324.79& 13.03(05)& 13.07(06)& 13.03(04)& 12.77(10)&   \nodata&Carbon Star\\

\enddata
\tablecomments{ \ Magnitude uncertainties are quoted in
  parentheses. For example, a magnitude of 11.75 $\pm$ 0.06 is listed
  as 11.75(06).}
\tablenotetext{a}{ \ To save space, the Source ID is truncated to show only the
  source coordinates.  The full electronic catalog (available through
  CDS) includes the {\it Spitzer} prefix (SSTOCEN).}
\tablenotetext{b}{ \ Fehrenbach's Star \citep{fehrenbach62}.}
\end{deluxetable}

\clearpage

%Table 6
\begin{deluxetable}{cccccc}
\tablewidth{0pc}
\tabletypesize{\small}
\tablecolumns{5}
\tablecaption{{\sc Sample Optical Depths Corresponding to [8] -- [24] =
  0.25 and 1.7} \label{tab:tau}}
\tablehead{&&\multicolumn{2}{c}{Silicate Dust} &
  \multicolumn{2}{c}{AlO$_{\rm x}$ Dust} \\ Optical Depth& [8] -- [24]
  &\colhead{$T$ =
    3297~K}&\colhead{$T$ = 3850~K} & \colhead{$T$ =
    3297~K}&\colhead{$T$ = 3850~K}}
\startdata

$\tau_{1}$ & 0.25 & 0.013 & 0.001 & 0.030 & 0.021 \\

$\tau_{2}$ & 1.70 & 0.120 & 0.056 & 0.220 & 0.090 \\

\enddata
\tablecomments{ \ $\tau_{1}$ and $\tau_{2}$ are examples of optical
  depth from \citet{groenewegen06} that correspond to [8] -- [24] colors of 0.25 and 1.7. These colors are marked in Figure~\ref{fig:indiv}.}
\end{deluxetable}

\clearpage

%Table 7
\begin{deluxetable}{llcccc}
\tablewidth{0pc}
\tabletypesize{\small}
\tablecolumns{6}
\tablecaption{{\sc Mass-Loss Rates of the Three Brightest M-type Stars}\label{tab:mlr}}
\tablehead{&&\multicolumn{4}{c}{$\dot{M}$ ($M_\odot$ yr$^{-1}$)}\\ &&\multicolumn{2}{c}{Silicate Dust} &
  \multicolumn{2}{c}{AlO$_{\rm x}$ Dust} \\ \colhead{LEID} &
  \colhead{Source ID}& \colhead{\textit{T} =
    3297~K}&\colhead{$T$ = 3850~K} & \colhead{$T$ =
    3297~K}&\colhead{$T$ = 3850~K}}
\startdata

33062&SSTOCEN J132630.19$-$472427.89 & 1.7 $\times$ 10$^{-7}$ & 1.3 $\times$ 10$^{-7}$ & 2.1 $\times$ 10$^{-7}$ & 1.5 $\times$ 10$^{-7}$ \\
44262&SSTOCEN J132646.37$-$472930.35 & 3.4 $\times$ 10$^{-8}$ & 2.8 $\times$ 10$^{-8}$ & 4.4 $\times$ 10$^{-8}$ & 3.5 $\times$ 10$^{-8}$ \\
35250&SSTOCEN J132737.72$-$472517.37 & 2.3 $\times$ 10$^{-8}$ & 1.8 $\times$ 10$^{-8}$ & 2.8 $\times$ 10$^{-8}$ & 2.2 $\times$ 10$^{-8}$ \\
&Total $\dot{M}$ for all dusty members & 3.4 $\times$ 10$^{-7}$ & 2.9 $\times$ 10$^{-7}$ & 4.2 $\times$ 10$^{-7}$ & 3.4 $\times$ 10$^{-7}$ \\

\enddata

\end{deluxetable}

\clearpage

%Table 8
\begin{deluxetable}{ccrrrrrr}
\tablewidth{0pc}
\tabletypesize{\tiny}
\tablecolumns{8}
%\rotate
\tablecaption{{\sc Fluxes of 70~\micron{} Sources} \label{tab:70flux}}
\tablehead{& & \multicolumn{6}{c}{Flux (mJy)} \\ \colhead{Source \#} &
  \colhead{Source ID} & \colhead{3.6~\micron{}} & \colhead{4.5~\micron{}} & \colhead{5.8~\micron{}} & \colhead{8~\micron{}} & \colhead{24~\micron{}} & \colhead{70~\micron{}}}
\startdata

  1&SSTOCEN J132723.86$-$472130.00 &15.94(0.32)& 10.23(0.20)&  6.66(0.13)&  5.59(0.11)&  2.42(0.05)& 75.79(17.31)\\
  2&SSTOCEN J132521.01$-$472623.43 & 1.96(0.04)&  1.49(0.03)&  1.96(0.04)& 12.33(0.25)& 14.33(0.29)&226.70(8.18)\\
  3&SSTOCEN J132523.57$-$472813.61 & \nodata   &  0.17(0.00)&  0.16(0.00)&  0.79(0.02)&  2.82(0.06)&246.73(18.94)\\
  4&SSTOCEN J132615.23$-$473024.56 &46.46(0.93)& 30.02(0.60)& 19.52(0.39)& 11.56(0.23)& 10.74(0.47)&158.02(7.96)\\
  5&SSTOCEN J132426.26$-$473905.88 & \nodata   &  \nodata   & \nodata    & \nodata    &  0.74(0.05)& 82.01(9.44)\\
  6&SSTOCEN J132714.06$-$474112.83 & 0.93(0.02)&  0.63(0.01)&  0.60(0.01)&  2.60(0.05)&  3.54(0.07)& 53.04(11.19)\\
  7&SSTOCEN J132655.87$-$474445.02 & 0.12(0.01)&  0.11(0.01)&  0.12(0.01)&  0.69(0.01)&  3.81(0.08)&131.46(23.81)\\
  8&SSTOCEN J132712.89$-$474739.59 & 0.07(0.01)&  0.05(0.01)&  0.05(0.01)&  0.06(0.01)&  0.75(0.02)&107.50(23.54)\\
  9\tablenotemark{a}&2MASX J13272621$-$4746042& 2.92(0.07)&  3.38(0.09)&  3.06(0.08)&  9.27(0.21)&  5.20(0.21)&617.08(73.84)\\

\enddata
\tablecomments{ \ Fluxes for 3.6 -- 24~\micron{} were determined
by PRF fitting with DAOphot and 70~\micron{} fluxes were determined
by aperture photometry with IRAF. Flux uncertainties are quoted in
  parentheses. For example, a flux of 15.94 $\pm$ 0.32 mJy is listed
  as 15.94(0.32).}
\tablenotetext{a}{ \ Source 9 is a resolved spiral
  galaxy (see Fig.~\ref{fig:gal}), while sources 1 through 8 are point
  sources.}
\end{deluxetable}

\clearpage

%Table 9
\begin{deluxetable}{ccrll}
\tablewidth{0pc}
\tabletypesize{\small}
\tablecolumns{5}
\tablecaption{{\sc Fitting Results for 70~\micron{} Sources} \label{tab:bb70}}
\tablehead{\colhead{Source \#} & \colhead{Source ID} &
  \colhead{$\chi^2/\rm dof$} & \colhead{$T_1$ (K)} & \colhead{$T_2$ (K)}}
\startdata

1 & SSTOCEN J132723.86$-$472130.00&35.8 & 4664 $\pm$ 19 & 58.5 $\pm$ 1.7 \\
2 & SSTOCEN J132521.01$-$472623.43& 3.3 & 2688 $\pm$ 25 & 67.5 $\pm$ 0.5 \\
3 & SSTOCEN J132523.57$-$472813.61&199.7& 1100 $\pm$ 200\tablenotemark{a}& 55.0 $\pm$ 10\tablenotemark{a} \\
4 & SSTOCEN J132615.23$-$473024.56&40.1 & 4329 $\pm$ 16 & 66.5 $\pm$ 0.8 \\
5 & SSTOCEN J132426.26$-$473905.88&75.0 & \nodata       & 50.0 $\pm$ 10\tablenotemark{a}  \\
6 & SSTOCEN J132714.06$-$474112.83&15.4 & 2901 $\pm$ 28 & 68.0 $\pm$ 2.0 \\
7 & SSTOCEN J132655.87$-$474445.02&52.2 & 1800 $\pm$ 200\tablenotemark{a}& 60.0 $\pm$ 10\tablenotemark{a} \\
8 & SSTOCEN J132712.89$-$474739.59&15.7 & 4000 $\pm$ 200\tablenotemark{a}& 50.0 $\pm$ 10\tablenotemark{a} \\
9\tablenotemark{b} & 2MASX J13272621$-$4746042&2.5  & 1084 $\pm$ 29 & 48.8 $\pm$ 0.8 \\

\enddata
\tablenotetext{a}{ \ The marked uncertainties represent twice the
  temperature increment in the range of fixed temperatures used for
  the underconstrained fits.}
\tablenotetext{b}{ \ Source 9 is a resolved spiral
  galaxy (see Fig.~\ref{fig:gal}), while sources 1 through 8 are point
  sources.}
\end{deluxetable}

\end{document}